\documentclass[ reprint,
superscriptaddress,
 amsmath,amssymb,
 aps,
prx, longbibliography
]{revtex4-1}
\usepackage{graphicx}
\usepackage{amsmath}
\usepackage{placeins}
\usepackage{color}
\usepackage{bm}

\newcommand{\Tc}{80}
\newcommand{\Pb}{Pb$_2$CaOsO$_6${}}
\newcommand{\PbCa}{Pb$_2$CaOsO$_6${}}
\newcommand{\PbCo}{Pb$_2$CoOsO$_6${}}
\newcommand{\PbZn}{Pb$_2$ZnOsO$_6${}}

\newcommand{\kkk}{\mathbf{k}}

\newcommand{\aaa}{\mathbf{a}}
\newcommand{\bbb}{\mathbf{b}}
\newcommand{\ccc}{\mathbf{c}}

\newcount\colveccount
\newcommand*\colvec[1]{
        \global\colveccount#1
        \begin{pmatrix}
        \colvecnext
}
\def\colvecnext#1{
        #1
        \global\advance\colveccount-1
        \ifnum\colveccount>0
                \\
                \expandafter\colvecnext
        \else
                \end{pmatrix}
        \fi
}

\begin{document}
\title{Magnetically induced metal--insulator transition in \Pb{}}

\author{Henrik Jacobsen}
\altaffiliation{Present address: Laboratory for Neutron Scattering and Imaging, Paul Scherrer Institute, CH-5232 Villigen PSI, Switzerland}
\email{henrik.jacobsen.fys@gmail.com}
\affiliation{Department of Physics, Oxford University, Clarendon Laboratory, Oxford, OX1 3PU, United Kingdom}

\author{Hai L. Feng} \email{hai.feng@iphy.ac.cn}
\affiliation{National Institute for Materials Science, 1-1 Namiki, Tsukuba, Ibaraki 305-0044, Japan}
\affiliation{Beijing National Laboratory for Condensed Matter Physics, Institute of Physics, Chinese Academy of Sciences, Beijing 100190, China}

\author{Andrew J. Princep}
\affiliation{Department of Physics, Oxford University, Clarendon Laboratory, Oxford, OX1 3PU, United Kingdom}
\affiliation{ISIS Pulsed Neutron and Muon Source, STFC Rutherford Appleton Laboratory, Harwell Campus, Didcot, OX11 0QX, United Kingdom}

\author{Marein C. Rahn}
\altaffiliation{Present address:  Institute for Solid State and Materials Physics, Technical University of Dresden, 01062 Dresden, Germany}
\affiliation{Department of Physics, Oxford University, Clarendon Laboratory, Oxford, OX1 3PU, United Kingdom}

\author{Yanfeng Guo}
\affiliation{School of Physical Science and Technology, ShanghaiTech University, Shanghai 201210, China} 
\author{Jie Chen} \affiliation{National Institute for Materials Science, 1-1 Namiki, Tsukuba, Ibaraki 305-0044, Japan}

\author{Yoshitaka Matsushita}
\affiliation{National Institute for Materials Science, 1-1 Namiki, Tsukuba, Ibaraki 305-0044, Japan}

\author{Yoshihiro Tsujimoto}
\affiliation{International Center for Materials Nanoarchitectonics (WPI-MANA), National Institute for Materials Science, 1-1 Namiki, Tsukuba, Ibaraki 305-0044, Japan}

\author{Masahiro Nagao}
\affiliation{Institute of Materials and Systems for Sustainability, Nagoya University, Nagoya 464-8601, Japan}

\author{Dmitry Khalyavin}
\affiliation{ISIS Pulsed Neutron and Muon Source, STFC Rutherford Appleton Laboratory, Harwell Campus, Didcot, OX11 0QX, United Kingdom}

\author{Pascal Manuel}
\affiliation{ISIS Pulsed Neutron and Muon Source, STFC Rutherford Appleton Laboratory, Harwell Campus, Didcot, OX11 0QX, United Kingdom}

\author{Claire A. Murray} 
\affiliation{Diamond Light Source Ltd, Harwell Science and Innovation Campus, Didcot, Oxfordshire OX11 0DE, United Kingdom}

\author{Christian Donnerer}
\affiliation{London Centre for Nanotechnology and Department of Physics and Astronomy, University College London, London WC1E 6BT, United Kingdom}

\author{James G. Vale}
\affiliation{London Centre for Nanotechnology and Department ofPhysics and Astronomy, University College London, London WC1E 6BT, United Kingdom}

\author{Marco Moretti Sala}
\altaffiliation{Present address: Dipartimento di Fisica, Politecnico di Milano, Piazza Leonardo da Vinci 32, I-20133 Milano, Italy}
\affiliation{European Synchrotron Radiation Facility, 71 Avenue des Martyrs, F-38000 Grenoble, France.}

\author{Kazunari Yamaura} \email{Yamaura.Kazunari@nims.go.jp}
\affiliation{International Center for Materials Nanoarchitectonics (WPI-MANA), National Institute for Materials Science, 1-1 Namiki, Tsukuba, Ibaraki 305-0044, Japan}

\author{ Andrew T. Boothroyd }
\email{andrew.boothroyd@physics.ox.ac.uk}
\affiliation{Department of Physics, Oxford University, Clarendon Laboratory, Oxford, OX1 3PU, United Kingdom}

\begin{abstract}
We report on the structural, magnetic, and electronic properties of two new double-perovskites synthesized under high pressure; \Pb{} and \PbZn{}.  Upon cooling below \Tc{}\,K, \Pb{} simultaneously undergoes a metal--insulator transition and develops  antiferromagnetic order. \PbZn{}, on the other hand, remains a paramagnetic metal down to 2\,K. The key difference between the two compounds lies in their crystal structure. The Os atoms in \PbZn{} are arranged on an approximately face-centred cubic lattice with strong antiferromagnetic nearest-neighbor exchange couplings. The geometrical frustration inherent to this lattice prevents magnetic order from forming down to the lowest temperatures. In contrast, the unit cell of \Pb{} is heavily distorted up to at least 500\,K, including antiferroelectric-like displacements of the Pb and O atoms despite metallic conductivity above 80\,K.
This distortion relieves the magnetic frustration, facilitating magnetic order which in turn drives the metal--insulator transition. Our results suggest that the phase transition in \Pb{} is spin-driven, and could be a rare example of a Slater transition.
\end{abstract}
\maketitle

\section{Introduction}

Metal--insulator transitions (MIT) can occur through a variety of mechanisms \cite{Imada1998,Kim2018a}.  Perhaps the best known is that of the Mott insulator, in which strong Coulomb interactions between electrons open up a gap at the Fermi energy \cite{Mott1968}. In the case of weak Coulomb interactions, Slater proposed that a MIT can be driven by magnetism alone: long-range antiferromagnetic order changes the periodicity of the crystal potential, splitting the electronic bands and potentially leading to a MIT \cite{Slater1951}. Such magnetically driven MITs are rare in real materials. 

Over the last two decades, a number of $5d$ transition-metal oxides have been suggested to undergo Slater transitions, including the perovskite osmate NaOsO$_3$ (Refs.~\onlinecite{Shi2009a,Calder2012}), and the iridate and osmate pyrochlores, $A_2$Ir$_2$O$_7$, ($A = $ Y or trivalent lanthanide), (Ref.~\onlinecite{Matsuhira2011}) and Cd$_2$Os$_2$O$_7$ (Ref.~\onlinecite{Mandrus2001}). The extended $5d$ states in these systems are relatively weakly interacting, as required for the Slater mechanism, but the band splitting caused by strong spin--orbit coupling can bring about Mott-like behavior instead.  Indeed, later work indicated that mechanisms other than the Slater mechanism may be behind the MIT in the above-mentioned $5d$ oxides
\cite{Kim2016a,Yamaura2012,Kim2018a}.

Recently, we have synthesised under high pressure several new double perovskites containing Os in the 5$d^2$ electronic configuration.\cite{Yuan-InorgChem-2015,Feng-PRB-2016,Princep2019,Chen-InorgChem-2020} These materials exhibit a variety of unusual and interesting electronic and magnetic phases, such as the magnetically driven loss of centrosymmetry into a polar metal phase that occurs in metallic \PbCo{}.\cite{Princep2019} 

Here  we investigate two other members of this family, \Pb{} and \PbZn{}.  We present resistivity, susceptibility, heat capacity, neutron diffraction and resonant inelastic x-ray scattering (RIXS) experiments, and discuss how the results shed light on the intriguing behavior of these compounds.
We find that \Pb{} undergoes a MIT simultaneously with the onset of magnetic order, as shown in Fig.~\ref{fig:Fig1}, and present evidence that the Slater picture might apply to this transition. \PbZn{} remains metallic and paramagnetic down to 2\,K, suggesting that the MIT in \Pb{} is tied to the presence of magnetic order. 

\section{Experimental}
Both polycrystalline and single crystal samples of \Pb{} were synthesized by a solid-state reaction from powders of PbO$_2$ (99.9\%, High Purity Chemicals. Co., Ltd., Japan), Os (99.95\%, Heraeus Materials Technology, Germany), and CaO$_2$ (Lab made). The powders were thoroughly mixed in a stoichiometric ratio, followed by sealing in a Pt capsule. The procedures were conducted in an Ar-filled glove box. To generate an applied pressure of 6\,GPa, static and isotropic compression of the capsule was achieved in a belt-type high pressure apparatus at National Institute for Materials Science (Kobe Steel, Ltd., Japan). The capsule was then heated at 1500$^\circ$C for one hour to synthesize the polycrystalline sample and at 1600$^\circ$C for two hours to synthesize the single crystal samples. It was then quenched to ambient temperature in less than a minute, after which the pressure was released. We note that we have observed some variability in the MIT temperature between samples, possibility due to small differences in composition. In all cases examined the MIT and the magnetic transition coincide in temperature. All the results presented here are for samples in which the MIT and magnetic transition occur at 80\,K.

Polycrystalline \PbZn{} samples were synthesized using the same method, with ZnO  instead of CaO$_2$ and KClO$_4$ as an oxygen source. Excess KCl was removed by washing with distilled water.

Electrical resistivity ($\rho$) of a single crystal of \Pb{} and a pressed powder of \PbZn{} was measured with a DC gauge current of 0.1 mA by a four-point method in a physical properties measurement system (PPMS, Quantum Design, Inc.). Electrical contacts were applied with Pt wires and Ag paste. The specific heat capacity ($C_p$) was measured in the same apparatus  on polycrystalline samples by a thermal relaxation method at temperatures between 2\,K and 300\,K. 
The magnetic susceptibility ($\chi$) of polycrystalline samples was measured in a magnetic properties measurement system (MPMS, Quantum Design, Inc.). The measurements were conducted at field cooling (FC) and zero-field cooling (ZFC) conditions, in a temperature range between 2\,K and 390\,K, under an applied magnetic field of 10\,kOe. 

A number of elastic neutron and x-ray scattering experiments on powders and single crystals of \Pb{} and \PbZn{} were performed. Preliminary neutron diffraction experiments on \Pb{} were carried out on the powder diffractometers SPODI operated by FRM II at the Heinz Maier-Leibnitz Zentrum (MLZ), Germany \cite{Hoelzel2015}, and HRPD at the ISIS Neutron and Muon Facility, UK. \cite{HRPD_data} Subsequently, we obtained more complete data on the WISH diffractometer at ISIS\cite{Chapon2011,WISH_data}, and these are the results presented in this paper.   Finally, high temperature x-ray powder diffraction experiments were carried out on I11 at the Diamond Light Source, UK. 

In the neutron powder diffraction experiments on WISH, 4.3\,g of \Pb{} was loaded in a cylindrical Al sample holder and measured at multiple temperatures between 2 and 105\,K for 20 min per run. 0.5\,g of \PbZn{} was loaded into a cylindrical vanadium sample holder and was measured at multiple temperatures between 2 and 300\,K, with runs at 2, 150 and 300\,K measured
for 3 hours each, and the remaining temperatures measured for 20 minutes each. For the structure refinements presented below we used the data from banks $(2+9)$ ($2\theta\sim58^\circ$), $(3+8)$ ($2\theta\sim 90^\circ$) and $(5+6)$ ($2\theta\sim153^\circ$). The data were reduced using MANTID \cite{Arnold2014}.

In the x-ray powder diffraction experiment on I11 at the Diamond Light Source, a series of small single crystals were crushed and ground into a micron sized powder, and then coated onto the outside of a glass capillary. This was done in order to minimize absorption from lead and osmium. An incident energy of 25\,keV was used. The capillary was rotated and a cryostream system provided a nitrogen atmosphere for heating and cooling the sample. Scans were made at 100, 290, 350, 425, and 500\,K.

To further understand the magnetism of \Pb{} we have carried out resonant inelastic x-ray scattering (RIXS) experiments on a \Pb{} crystal at the Os $L_3$ edge ($E=10.877$\,keV, $2p\rightarrow 5d$) at 20\,K on the ID20 spectrometer at the European Synchrotron Radiation Facility, Grenoble \cite{MorettiSala2018}. We measured the RIXS spectrum at a wavevector of ${\bf Q}=(-2.5,7,0)$ (which corresponds to $(0,7,2.5)$ when indexed on the undistorted cell). We used a relaxed resolution setting ($\Delta E = 0.40(1)$\,eV, full width at half maximum) to measure the spectrum up to an energy loss of 11\,eV, and a high resolution mode ($\Delta E = 0.064(1)$\,eV) to probe the low energy excitations.

\section{Results}
\begin{figure}
\includegraphics[width=0.35\textwidth]{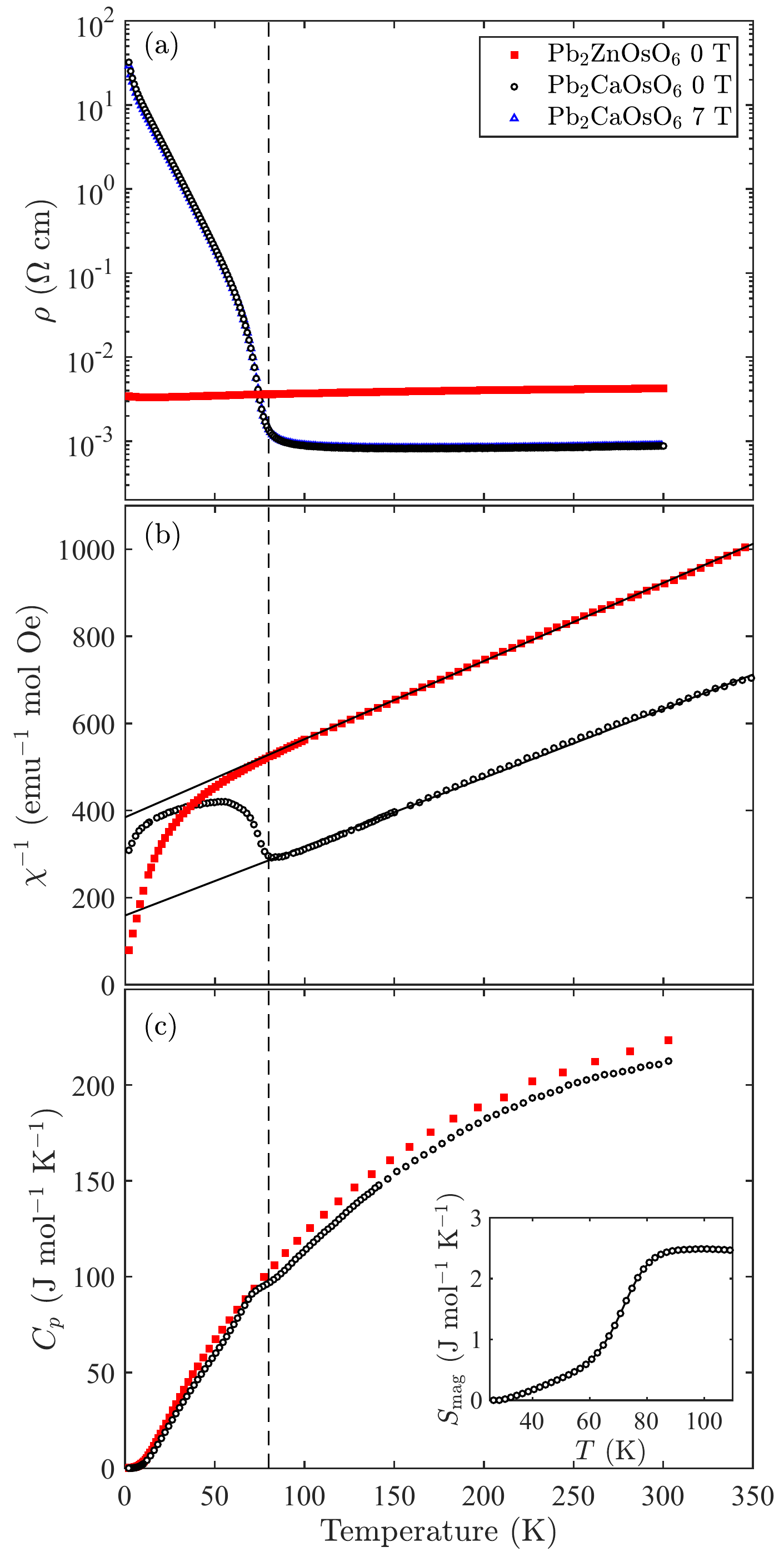}
\caption{ Bulk measurements on \Pb{} and \PbZn{} as a function of temperature. (a) Resistivity, showing  that \Pb{} undergoes a MIT at 80\,K, whereas \PbZn{} remains metallic down to 2\,K. The resistivity curve for \Pb{} is unaffected by an external magnetic field of 7\,T.
(b) Inverse susceptibility together with a fit to the Curie--Weiss law at high temperature. At 80\,K there is an anomaly in \Pb{}, and at approximately the same temperature \PbZn{} starts to deviate from the Curie--Weiss law.
(c) Heat capacity showing an anomaly at around 80\,K in \Pb{}. The inset shows the magnetic entropy of \Pb{} calculated by subtracting a polynomial approximation of the lattice contribution to the heat capacity. 
}
\label{fig:Fig1}
\end{figure}

Figure~\ref{fig:Fig1}(a) shows the resistivity of a single crystal of \Pb{} and a pressed powder of \PbZn{}. At room temperature the resistivity is $\rho\sim 8\times 10^{-4}$\,$\Omega$\,cm for \Pb{} and  $\sim 4\times 10^{-3}$\,$\Omega$\,cm for \PbZn{}. Upon cooling, the resistivity of both compounds decreases linearly with temperature, consistent with metallic conduction. At $T_{\rm c} \simeq \Tc{}$\,K, \Pb{} undergoes a transition from a metal to an insulator, with its resistivity increasing by more than four orders of magnitude between 80\,K and 2\,K. The resistivity of \Pb{} is unchanged by the application of a 7\,T magnetic field. 

Figure~\ref{fig:Fig1}(b) shows the inverse susceptibility measured on powder samples. 
At high temperatures, the data for both materials follow the Curie--Weiss law $\chi = C/(T+\theta)$ with $\theta=100.7(3)$\,K and effective moment $\mu_\text{eff}=2.25(1)$\,$\mu_{\rm B}$/Os
for \Pb{}, and $\theta=214.4(4)$\,K and $\mu_\text{eff}=2.11$\,$\mu_{\rm B}$/Os for \PbZn{}. 
In the absence of spin--orbit coupling we would expect localized $5d^2$ electrons of Os$^{6+}$ to form a $S=1$ state with an effective moment of $\mu_\text{eff} = 2\sqrt{2} = 2.83$\,$\mu_\text{B}$. The reduction of 20--25\% from the spin-only value indicates the influence of spin--orbit coupling, as expected for $5d$ orbitals\cite{Morrow2016}.

An anomaly in the susceptibility at $T_{\rm N}\simeq \Tc$\,K signals the transition to an antiferromagnetically ordered state in \Pb{}. At roughly the same temperature the susceptibility of \PbZn{} deviates from the Curie--Weiss law, but no phase transition is observed. The simultaneous onset of the MIT and magnetic ordering ($T_{\rm c}= T_{\rm N}$ within experimental resolution) in \Pb{} indicates that the two phenomena are related.

The temperature dependence of the heat capacity of both compounds is presented in Fig.~\ref{fig:Fig1}(c). The data for \Pb{} has an anomaly close to $T_{\rm c}$. The inset displays the magnetic entropy, calculated after subtraction of a polynomial approximation to the phonon contribution to the heat capacity. The entropy associated with this transition is $S\approx 0.3R$. 

\begin{figure}
    \centering
            \includegraphics[width=0.48\textwidth]{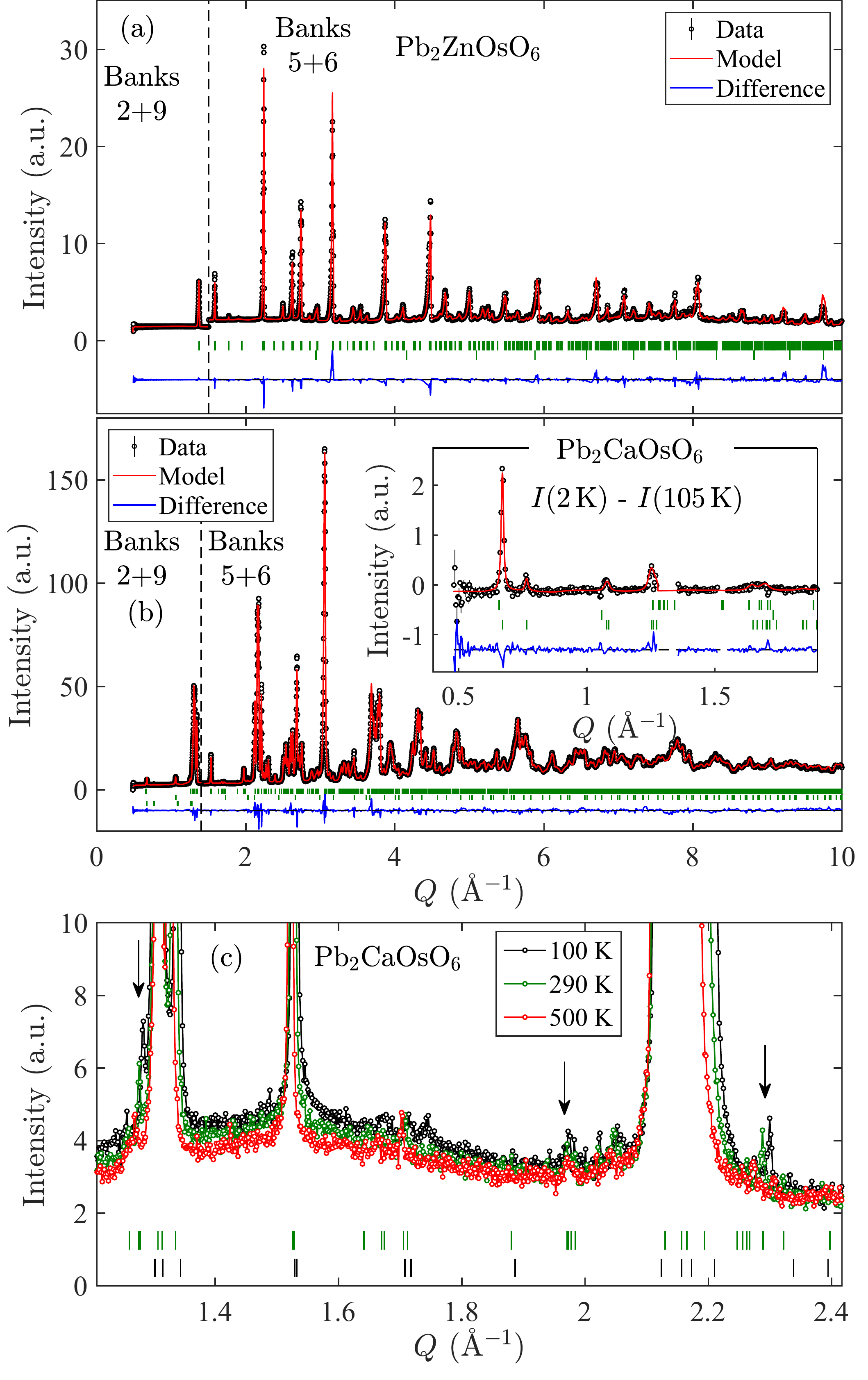}
\caption{Examples of our neutron diffraction data on (a) \PbZn{} and (b) \Pb{} at 2\,K. Data at low $Q$ (below dotted line) is from banks 2 and 9, data at higher $Q$ is from banks 5 and 6 of the WISH diffractometer. The inset in (b) shows the magnetic scattering found by subtracting the data at 105\,K from the low temperature data. In (a) the green lines mark   Bragg peaks from \PbZn{} and vanadium, in (b) the lines mark Bragg peaks from \Pb{}, the Pb$_2$Os$_2$O$_7$ impurity and the magnetic phase of \Pb{}. (c) Synchrotron X-ray powder diffraction data on \Pb{} at high temperatures. The arrows point to some of the reflections from the structural distortion. The green (black) vertical lines indicate Bragg peaks from the distorted (undistorted) cell. }
\label{fig:Fig2}
\end{figure}

We show examples of our neutron diffraction data at 2\,K on \PbZn{} and \Pb{} in Fig~\ref{fig:Fig2}(a) and (b), respectively.  The red line through the data shows our refinement of the crystal structure using the FullProf Suite \cite{Rodriguez-Carvajal1993}, and the tick marks indicate the positions of Bragg peaks. In addition to the \PbZn{} main phase we also account for the vanadium sample holder in Fig.~\ref{fig:Fig2}(a). In Fig.~\ref{fig:Fig2}(b) the tick marks indicate the \Pb{} main structural phase, the magnetic phase, and a few percent of Pb$_2$Os$_2$O$_7$ \cite{Badaud1972}. We have established that Pb$_2$Os$_2$O$_7$ remains paramagnetic at  temperatures down to 2\,K (see Appendix~A), so its presence has no effect on any of the results we present. We also observe small peaks from the Al sample holder and trace amounts of unreacted PbO$_2$, neither of which scatter strongly enough to warrant inclusion as separate phases in the refinement.

We find that \PbZn{} crystallizes in space group P$2_1$/n with lattice parameters $a=5.6329(2)$\,\AA{}, $b=5.6059(2)$\,\AA{}, $c=7.9201(2)$\,\AA{}, $\beta=89.96(1)^\circ$ at 2\,K, similar to \PbCo{} \cite{Princep2019} (see Table~\ref{tab:Tab2} in Appendix~B for a list of all the refined structural parameters). The monoclinic structure of \PbZn{} is typical for many double perovskites \cite{Howard2003} and combines octahedral tilting ($a^+b^-b^-$ tilting pattern in the Glazer notation \cite{Glazer1972}) and rock-salt ordering of Zn and Os in the B-site perovskite position. 

\begin{figure}
    \centering
    \includegraphics[width=0.22\textwidth]{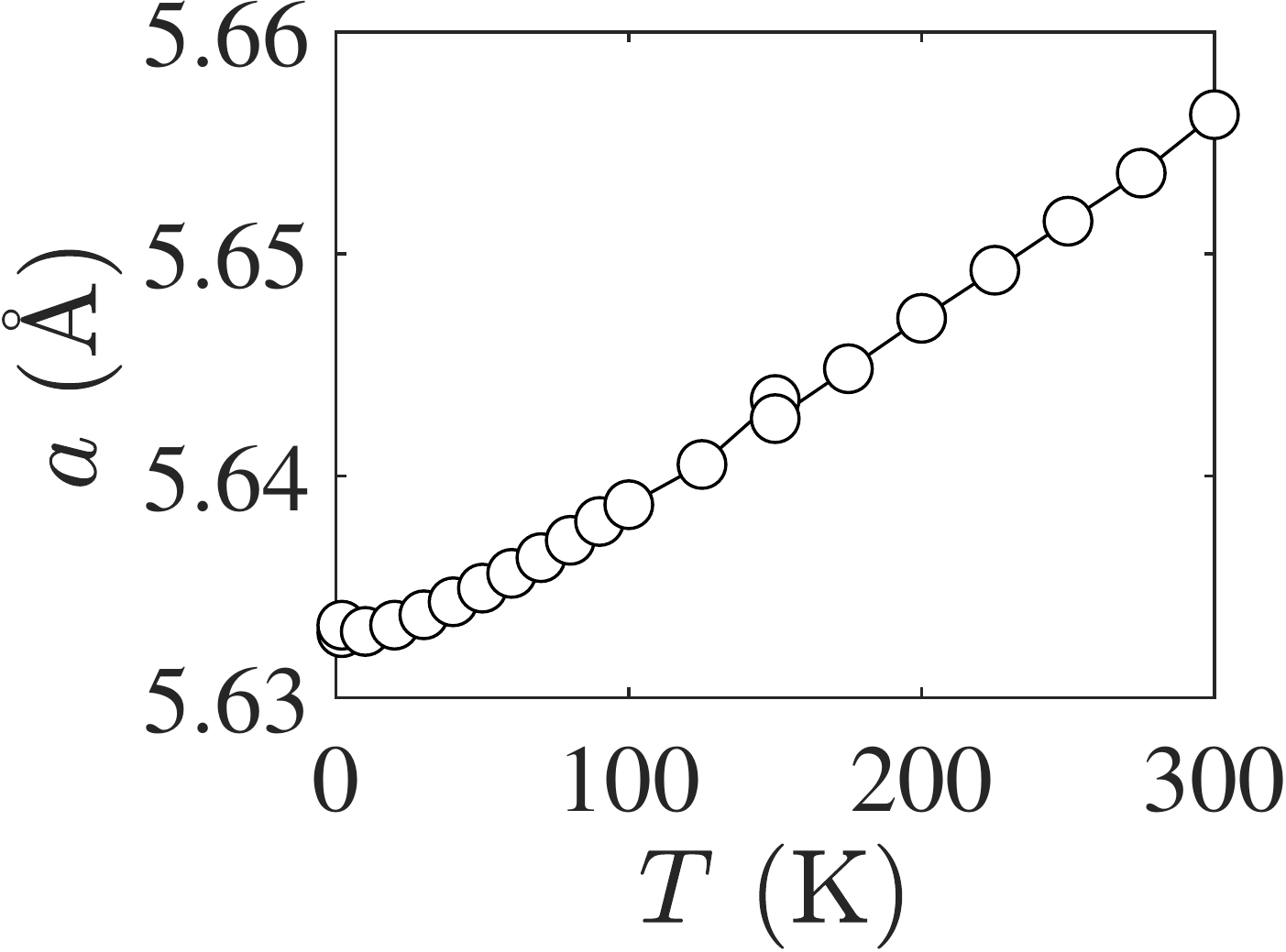}
    \includegraphics[width=0.22\textwidth]{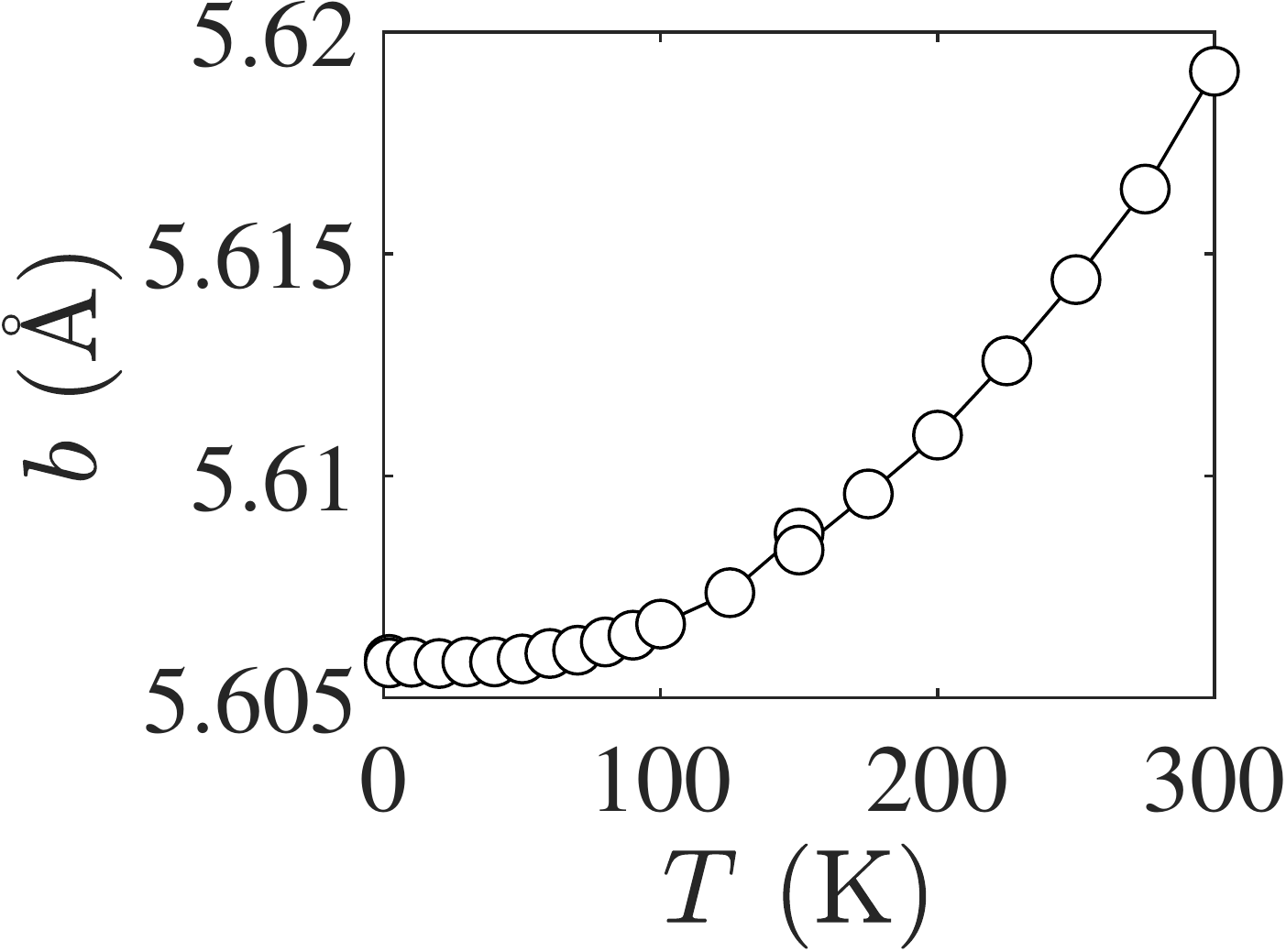}
    \includegraphics[width=0.22\textwidth]{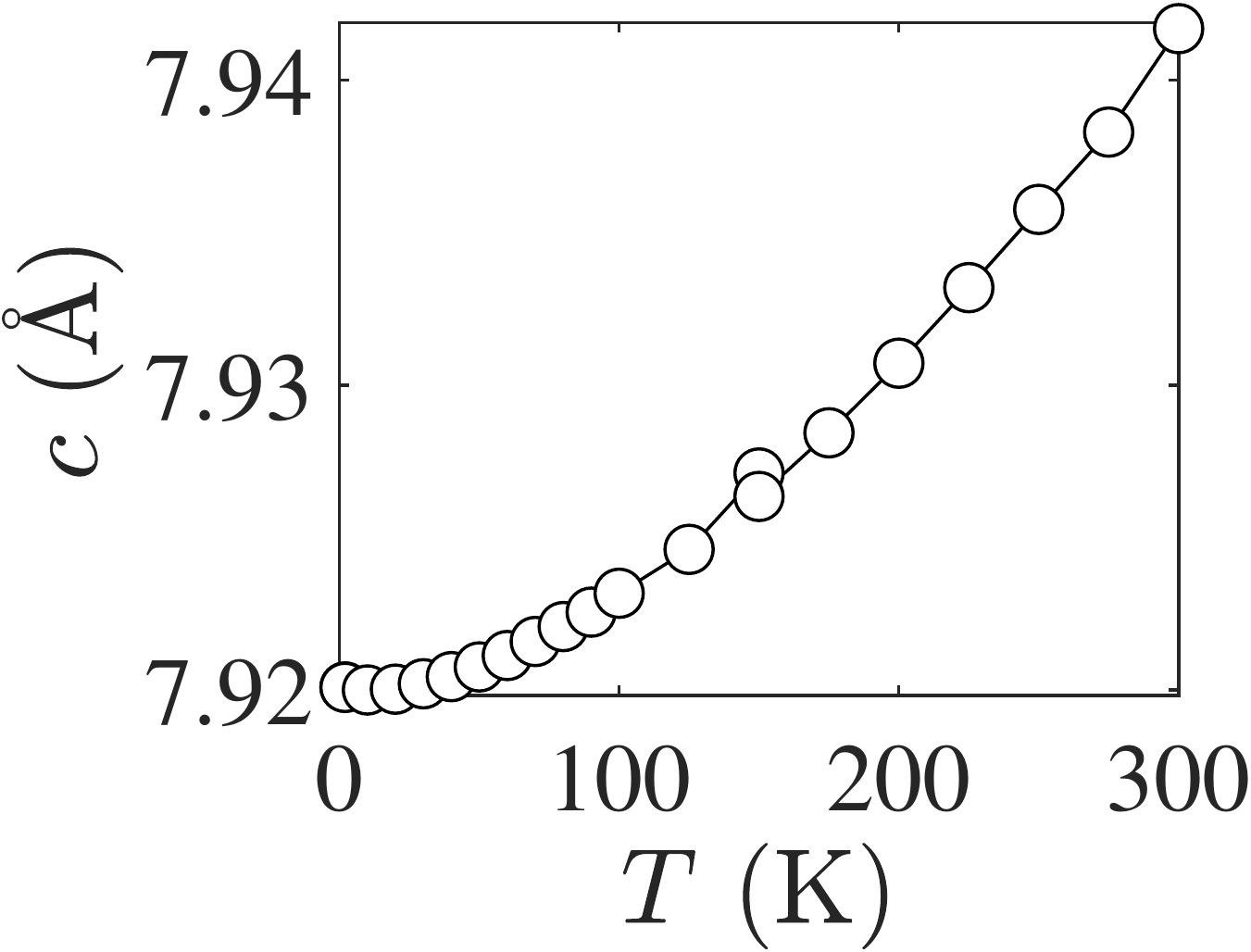}
    \caption{Temperature dependence of the lattice parameters for \PbZn{} determined from neutron powder diffraction measurements.  The statistical uncertainty on the parameters is smaller than the markers.}
    \label{fig:Fig3}
\end{figure}

\begin{figure}
    \centering
\includegraphics[width=0.48\textwidth]{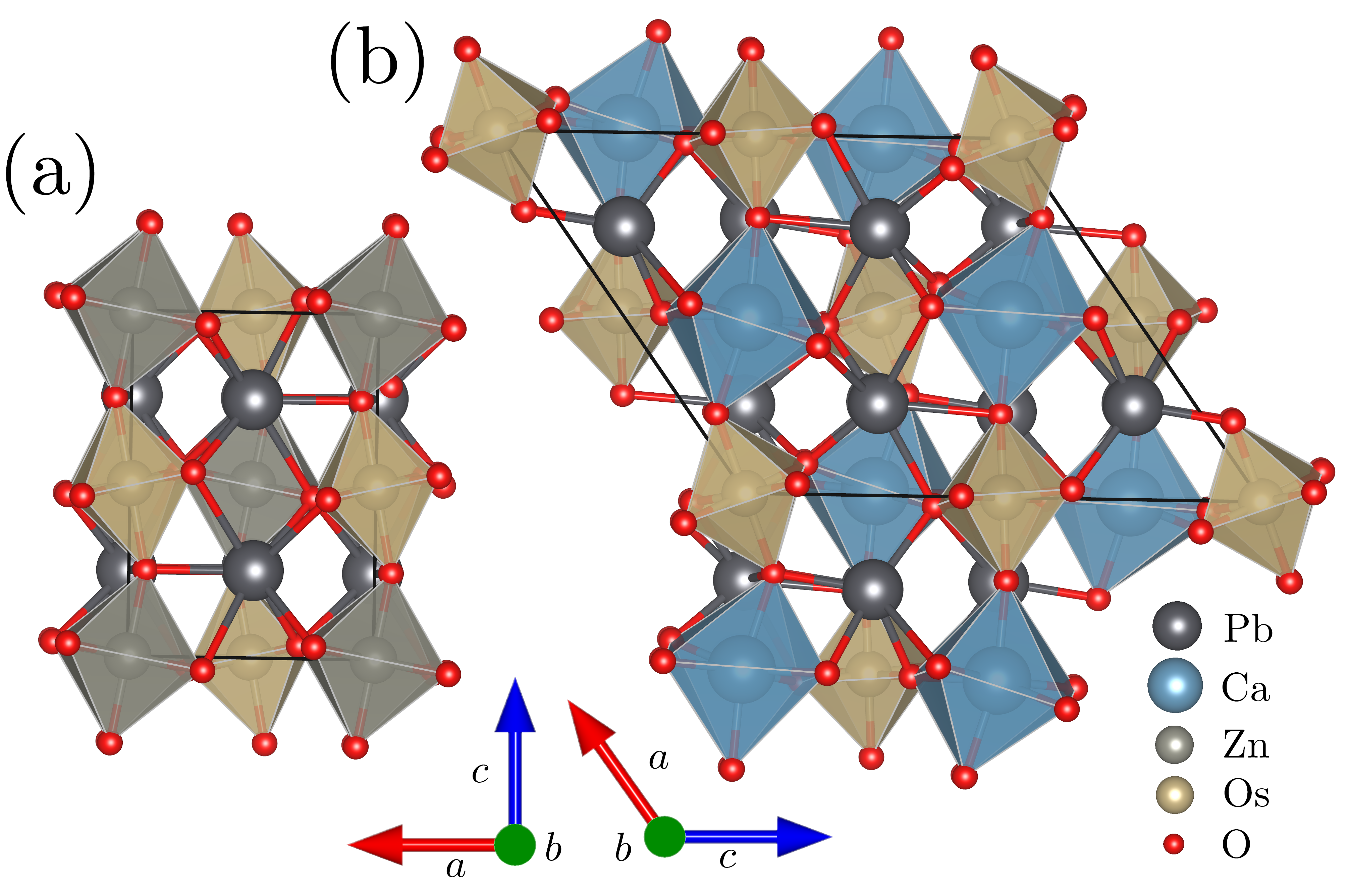}
    \caption{The unit cell of (a) \PbZn{} and (b) \Pb. The distortion in \Pb{} is mainly seen in the oxygen octahedra which are tilted and deformed. }
    \label{fig:Fig4}
\end{figure}

There is no evidence of short- or long-range magnetic order down to 2\,K. No anomalies are seen in the lattice parameters between 2 and 300\,K (see Fig.~\ref{fig:Fig3}), and we find no evidence for oxygen vacancies. Fig.~\ref{fig:Fig4}(a) shows the crystal structure of \PbZn{} as viewed down the $b$ axis.

\Pb{} also crystallizes in space group  P$2_1$/n, but in a significantly distorted cell with lattice parameters $a=10.0812(3)$\,\AA{}, $b=5.689(1)$\,\AA{}, $c=11.837(4)$\,\AA{}, $\beta=125.32(2)$ at 2\,K (see Table~\ref{tab:Tab2} in Appendix~B). The structure can be seen as a distortion of the \PbZn{} structure, with $a=5.920(1)$\,\AA{}, $b=5.688(1)$\,\AA{}, $c=8.22(1)$\,\AA{}, $\beta=90.61(1)$ at 2\,K. The distortion has propagation vector $\kkk=(\tfrac{1}{2},0,\tfrac{1}{2})$. The unit cell is shown in Fig.~\ref{fig:Fig4}(b).  Our synchrotron x-ray powder diffraction experiments, Fig.~\ref{fig:Fig2}(c), show that this structural modification is present at all measured temperatures up to 500\,K, which confirms that it is not directly related to the MIT at $T_{\rm c} = 80$\,K.

More insight into the structure of \Pb{} can be obtained by decomposing the structure in terms of symmetrized displacive modes of the parent cubic perovskite structure with Pm$\bar{3}$m symmetry. This allows us to identify the primary distortions and compare them with other perovskite structures. The result of the decomposition made with the ISODISTORT software \cite{isodistort,Campbell2006} is summarized in Appendix~D Table \ref{tab:Tab6} for \PbZn{} and Table \ref{tab:Tab6} for \Pb{}. 

Similarly to \PbZn{}, \Pb{}  contains the $R^{+}_{4}$ $({\bf k}=1/2,1/2,1/2)$  and $M_3$ $({\bf k}=1/2,1/2,0)$ irreps, which yield the $a^+b^-b^-$ Glazer tilt pattern. In \Pb{}, however, the largest distortion is $\Lambda_3$ $({\bf k}=1/4,1/4,1/4)$. The distortions associated with $\Lambda_3$ are less common but were also observed in several Bi$^{3+}$-containing perovskites such as BiCrO$_3$ \cite{Khalyavin2019}, BiScO$_3$ \cite{Khalyavin2014} and BiFe$_{1/2}$Sc$_{1/2}$O$_3$ \cite{Prosandeev2014}. We will discuss the significance of the $\Lambda_3$ distortion in the next section. 

The distortion is most pronounced in the oxygen atoms and causes the oxygen octahedra surrounding the Ca and Os sites to tilt and deform. The Os--O bond lengths at 2\,K are listed in Table~\ref{tab:Tab1} for both compounds. The bond directions are given with respect to the \PbZn{} unit cell. The oxygen octahedra in \PbZn{} are  elongated along $c$. Notably, in \Pb{} there are two different Os sites with different local surroundings. The octahedron surrounding Os1 is compressed along $(\aaa-\bbb)$, whereas the Os2 octahedron is close to ideal. There is also a significant distortion of the local coordination of the Pb atoms in the two compounds, as detailed in Appendix B.

\begin{table}[]
    \centering
    \begin{tabular}{lccc}
    \hline \hline
          & $ (\aaa-\bbb)$       & $ (\aaa+\bbb)$       & $ \ccc$  \\ \hline
         \PbZn     & 1.879(10) & 1.879(10) & 2.005(10) \\
         \Pb{} Os1 & 1.833(10) & 1.933(10) & 1.919(8) \\
         \Pb{} Os2 & 1.924(9)  & 1.979(10) & 1.984(11) \\ \hline \hline
    \end{tabular}
    \caption{The Os-O bond lengths (in \AA{}) in \PbZn{} and \Pb{} at 2\,K. The direction of the bond is given relative to the \PbZn{} unit cell. }
    \label{tab:Tab1}
\end{table}

\begin{figure}
    \centering
\includegraphics[width=0.48\textwidth]{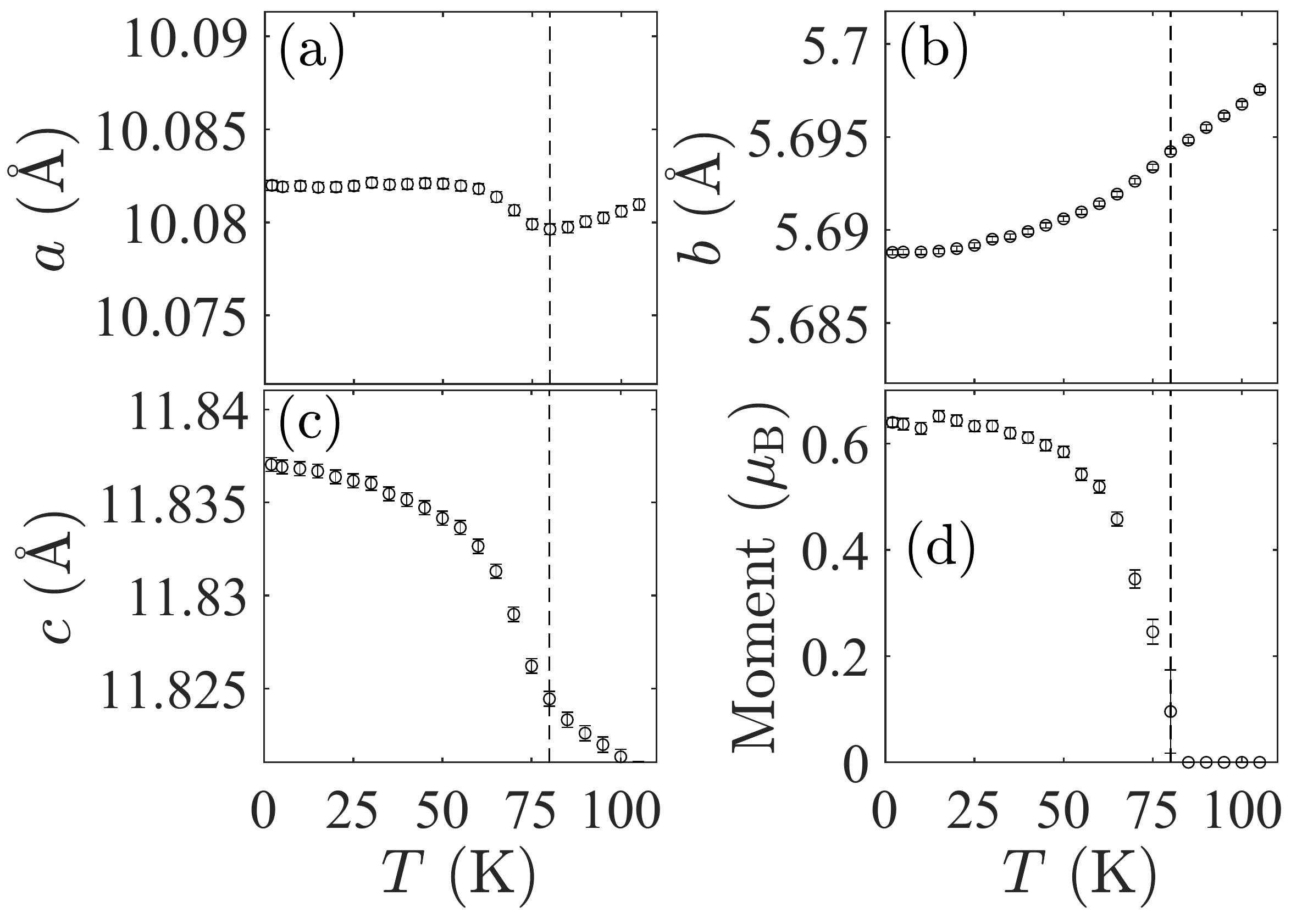}
    \caption{Temperature dependence of the lattice constants and ordered magnetic moment of \Pb{} determined from neutron powder diffraction measurements. The vertical line marks the metal--insulator and magnetic ordering transitions, which coincide at $T_{\rm c} = T_{\rm N} = 80$\,K. In (d) the moment is shown for the non-collinear model.}
    \label{fig:Fig5}
\end{figure}

The temperature dependence of the lattice parameters of \Pb{} for $T\leq105$ K is shown in Fig.~\ref{fig:Fig5}. The $a$ and $c$ parameters, and to a lesser extent the $b$ parameter, exhibit anomalies at $T_{\rm c}$. The fact that the lattice relaxes when magnetic order sets in indicates a significant magnetoelastic coupling, as also found in NaOsO$_3$ (Ref.~\onlinecite{Calder2012}).
Such an effect could be the result of the spatially extended $5d$ orbitals of Os bonding covalently with their environment. Importantly, we observe no symmetry changes in the crystal structure, and no abrupt changes in e.g.~the oxygen octahedra. We conclude from these results that no structural phase transition accompanies the MIT. 
We observe no further anomalies in the lattice parameters for temperatures above $T_{\rm c}$ up to 500\,K. 

At temperatures below $T_{\rm N}$ we observe the emergence of additional resolution-limited peaks in the neutron powder diffraction data on \Pb{}. The peak positions are consistent with magnetic order with propagation vector ${\bf k} = (\tfrac{1}{2},\tfrac{1}{2},0)$. The magnetic intensity, which is much weaker than the nuclear scattering, can be isolated by plotting the difference between the data recorded at 2\,K and 105\,K, as shown in the inset to Fig.~\ref{fig:Fig2}(b). The subtraction is imperfect due to the change in lattice constant around $T_{\rm N}$, but several magnetic peaks are clearly visible despite this complication, with $(\tfrac{1}{2},\tfrac{1}{2},0)$ being the strongest. 

\begin{figure}
    \centering
        \includegraphics[width=0.48\textwidth]{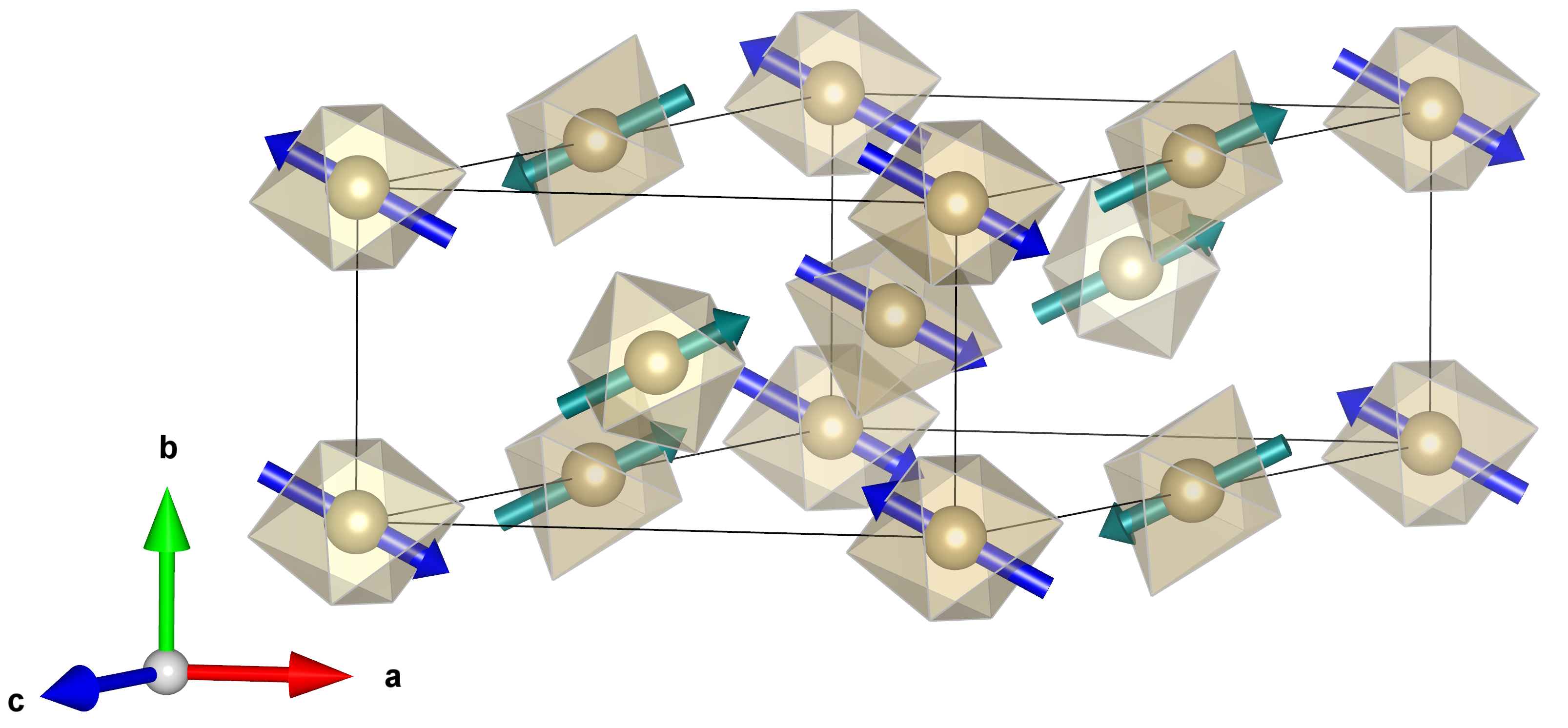}
    \caption{    The unit cell of \Pb{} with arrows indicating magnetic moments on the Os sites for the non-collinear model. The different colors of the arrows represent the two different Os atoms, with dark blue representing Os1 and teal representing Os2.  }
    \label{fig:Fig6}
\end{figure}

There are no symmetry constraints on the magnetic structure (the magnetic space group is P$_s\bar{1}$) (see Appendix~C for details). The absence of symmetry restrictions substantially complicates the analysis of the magnetic structure, making unambiguous determination of the spin configuration, based on the available powder diffraction data, impossible. To reduce the number of parameters in the refinement, we restricted our consideration to two cases: (i) when all spins are constrained to lie along a single direction (collinear model) and the moments on the symmetry-independent sites (Os1 and Os2) are allowed to adopt different magnitudes; (ii) when both Os1 and Os2 sites are constrained to have identical moment size but canting between the sublattices is allowed (non-collinear model). Both models can reproduce the observed magnetic intensities accurately and cannot be discriminated based on the quality of the agreement with the data. The collinear model was tested assuming that the magnetic moments are confined either within the $ac$-plane or along the $b$-axis and only the former provided a good fit. The refinement converged with substantially different moment sizes on Os1 and Os2, 0.87(1) $\mu_{\rm B}$ and 0.26(1) $\mu_{\rm B}$, respectively, and this difference is essential to achieve a good fit. The refinement in the non-collinear model converged with moment size 0.64(1) $\mu_{\rm B}$ and canting angle $56^\circ$ between the Os1 and Os2 spins. Fig.~\ref{fig:Fig6} illustrates the structure of the non-collinear model. 
The average ordered moment per site is similar for both models; 0.57(1)  $\mu_{\rm B}$ in the collinear model and  0.64(1)  $\mu_{\rm B}$ in the non-collinear model. The temperature dependence of the ordered  moment in the non-collinear model is plotted in Fig.~\ref{fig:Fig5}(d), and saturates at 0.64(1)\,$\mu_{\rm B}$/Os at 2\,K. 

\begin{figure}
\includegraphics[width=0.48\textwidth]{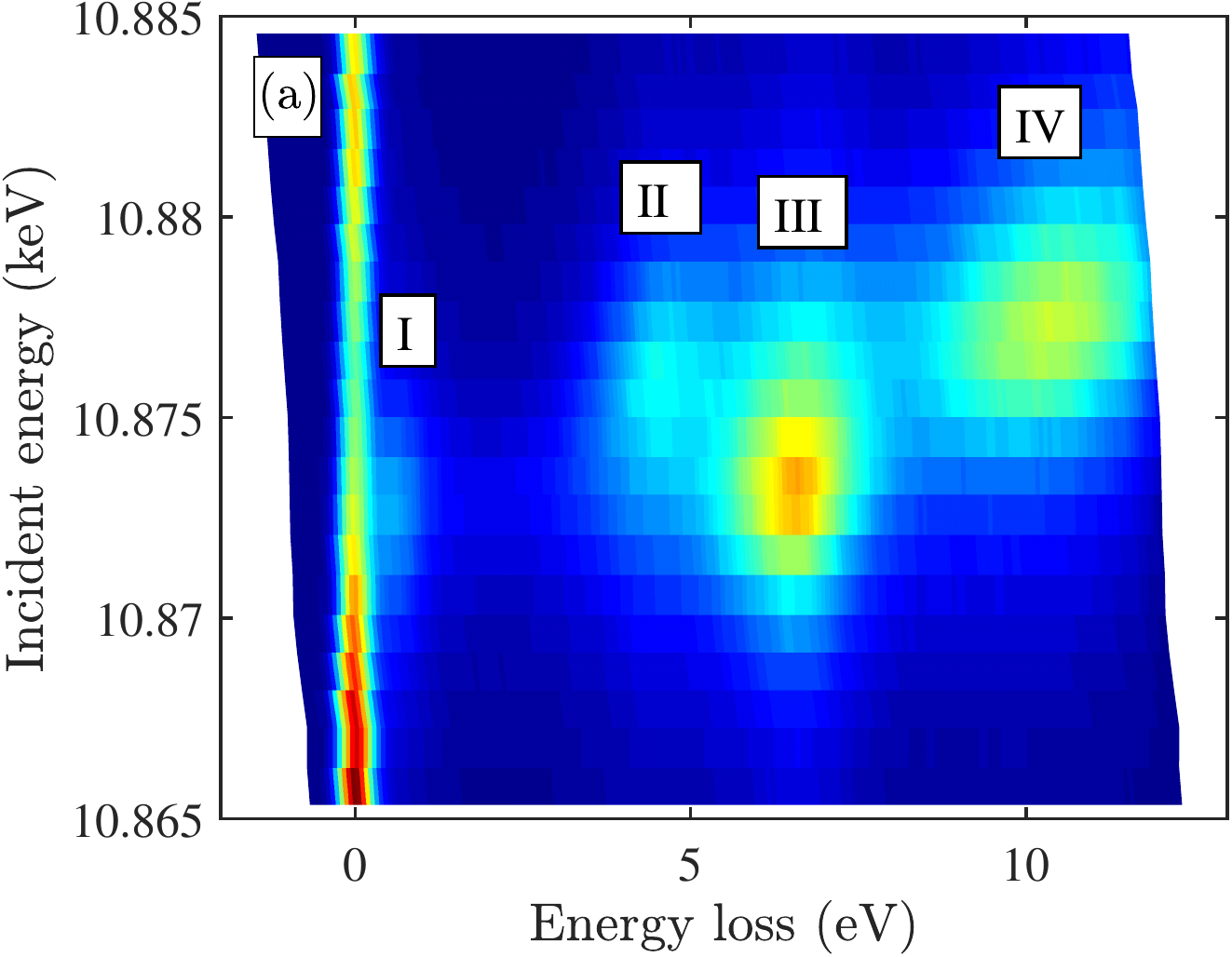}
\includegraphics[width=0.22\textwidth]{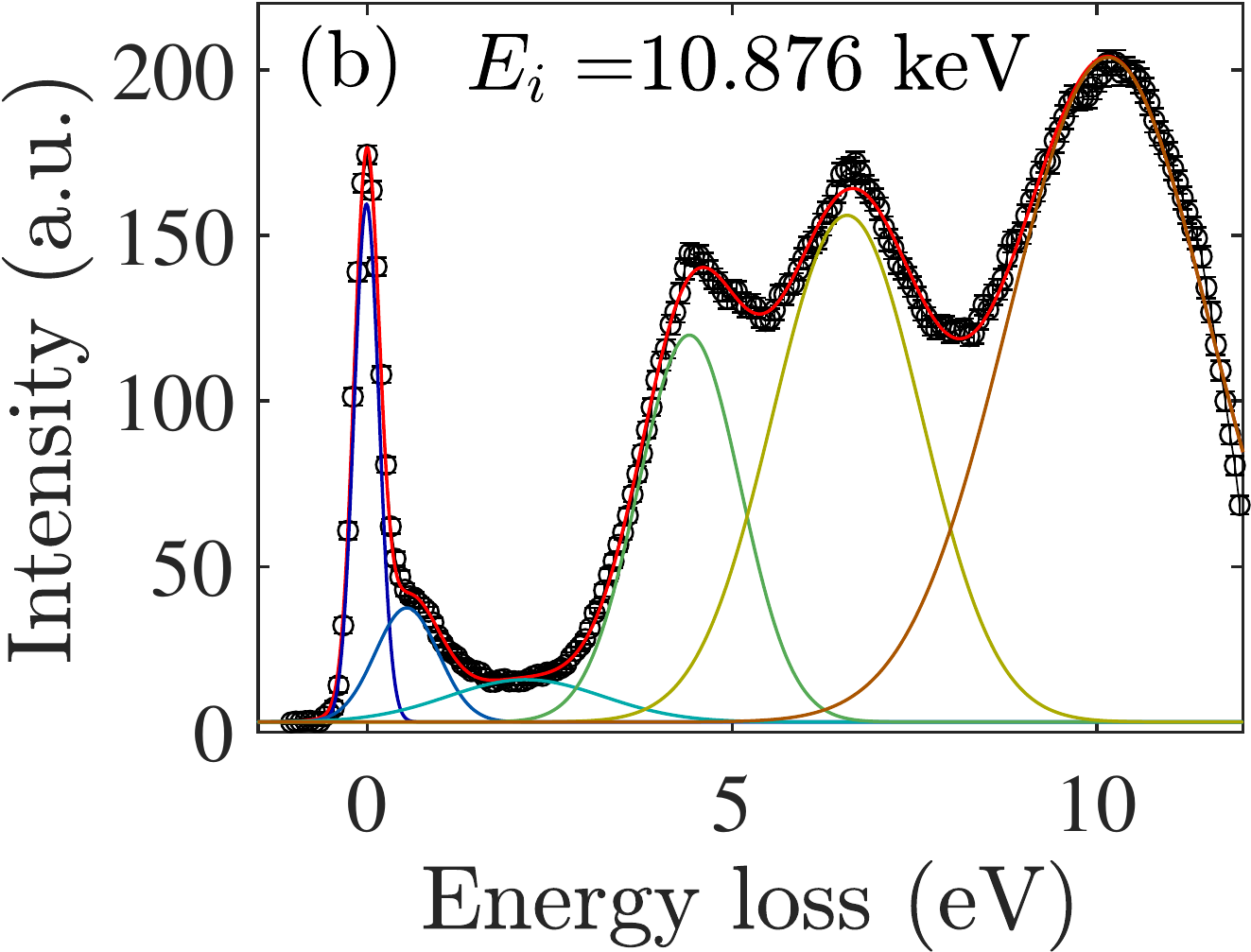}
\includegraphics[width=0.22\textwidth]{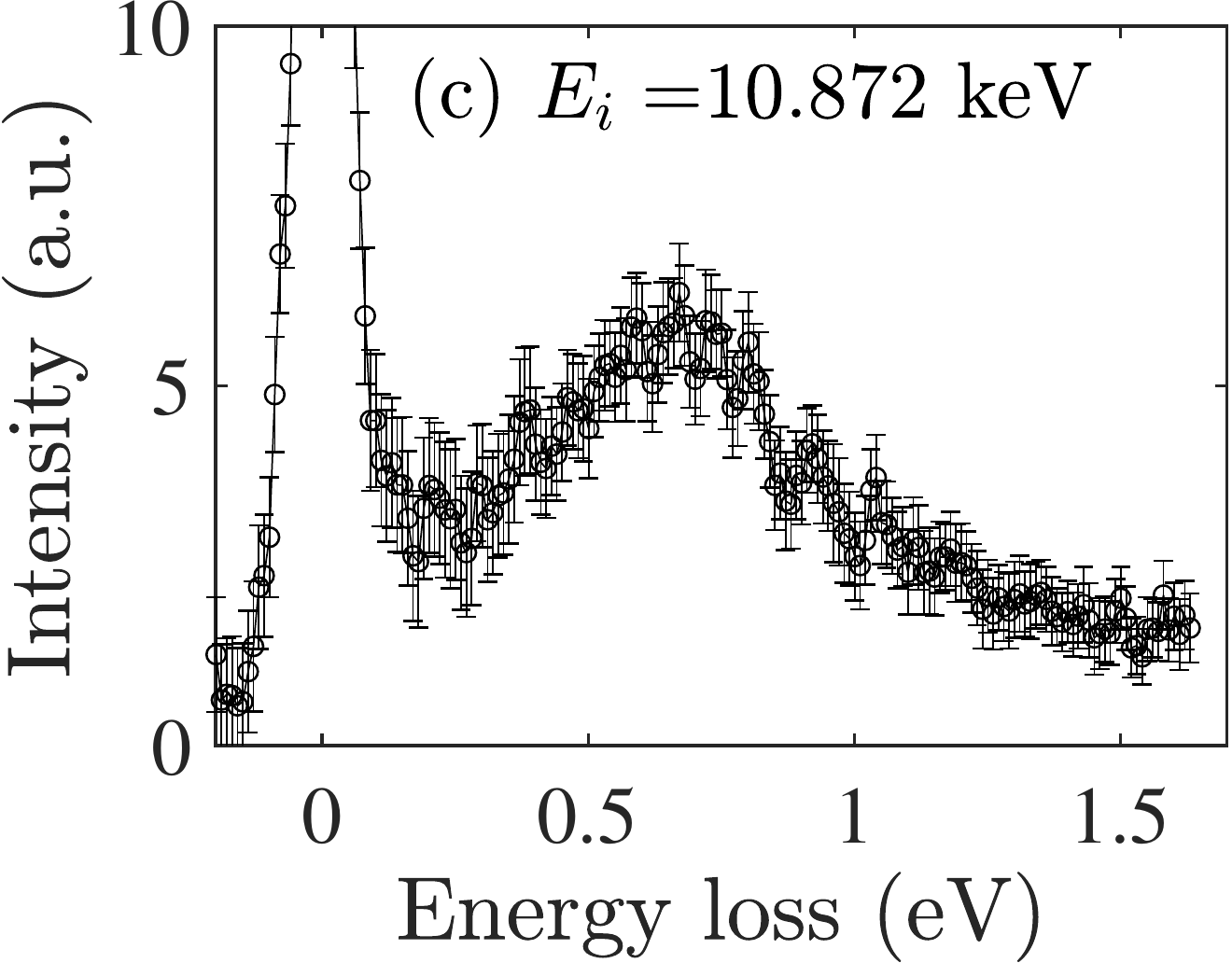}
\caption{RIXS data on \Pb{} at ${\bf Q}=(-2.5,7,0)$ at $T=20$ K. (a) shows a map with four distinct excitations. (b) A scan through the map at $E_{\rm i}=10.876$\,keV. (c) A high resolution scan through $E_{\rm i}=10.972$\,keV to look at the intra-$t_{2g}$ levels.}
\label{fig:Fig7}
\end{figure}

Information about the electronic ground state of Pb$_2$CaOsO$_6$ was sought from resonant inelastic x-ray scattering (RIXS). Figure~\ref{fig:Fig7}(a) shows a RIXS map measured with relaxed energy resolution. There are four features, labelled I to IV in the figure, which resonate at slightly different energies. Based on results from other osmates with Os in octahedral coordination \cite{Calder2016,Vale2018a}, these processes can be assigned as follows: I is below 1\,eV and corresponds to intra $t_{2g}$ excitations; II and III at around 4 and 7\,eV correspond to  $t_{2g}\rightarrow e_g$ transitions; IV corresponds to charge transfer excitations. Figure~\ref{fig:Fig7}(b) shows a cut through this data for incident energy $E_\text{i}=10.876$\,keV  which clearly displays the four broad features. 
The splitting of the $t_{2g}$ and $e_g$ levels by $10Dq\sim4$\,eV (excitation II) is similar to that in other osmates \cite{Calder2016}. 

In Fig.~\ref{fig:Fig7}(c) we show high resolution RIXS data recorded at $E_\text{i}=10.872$\,keV in order to further investigate the intra-$t_{2g}$ excitations. Even in the high resolution setting, we observe only a single broad peak centered around 0.7\,eV. The lack of distinct peaks precludes further analysis of the spectrum. The width of the peaks, which far exceeds the energy resolution, is an indication of the itinerant nature of the $5d$ electrons, as found also in NaOsO$_3$.\cite{Vale2018,Vale2018a}

\section{Discussion}

We first consider the crystal structures of \PbZn{} and \Pb{} in more detail. For \PbZn{}, the Goldschmidt tolerance factor, which describes the mismatch between the size of the A and B sites \cite{Goldschmidt1926} is $t=1.001$. The sizes of the Pb$^{2+}$, Zn$^{2+}$ and Os$^{6+}$ ions are thus almost perfectly matched for the double perovskite structure. Despite this, the oxygen octahedra in \PbZn{} are tilted (mainly by the $R_4^+$ and $M_3^+$ modes), and the space group is not cubic, but P2$_1$/n. This distortion is probably caused by the interaction between the Pb lone pair and the surrounding oxygen, which is known to cause structural distortions, for example in PbO (Ref.~\onlinecite{Walsh2005}) and Pb$_2$MnTeO$_6$ (Ref.~\onlinecite{Retuerto2016}).

The tolerance factor for \Pb{} is $t=0.941$, which indicates that the large ionic radius of Ca does not fit as well in the Pb$_2B$OsO$_6$ structure as the smaller Co or Zn ions. In \Pb{} the crystal structure is heavily distorted up to temperatures of at least 500\,K, and due to this distortion it is questionable if it can even be classified as a double perovskite. The structure of \Pb{} appears to be unique among double perovskites, as it is not shared by any of the 1,112 double perovskites surveyed in Ref.~\onlinecite{Vasala2015}. The structural distortion in \Pb{} is most likely caused by the combined efforts of the large Ca ionic size and the Pb lone pair. 

The lone pair electronic instability is well-known for many Pb$^{2+}$ -based perovskite systems and this instability is responsible for their ferroelectric (e.g.~PbTiO$_3$ and PbVO$_3$, Refs.~\onlinecite{Kupriyanov1998} and \onlinecite{Shpanchenko2004}) and antiferroelectric (e.g.~PbZrO$_3$ and Pb$_2$MnWO$_6$, Refs.~\onlinecite{Fujishita1982} and \onlinecite{Orlandi2014}) properties. Its effect is most clearly elucidated by analysing the displacement modes in \Pb{}. In addition to tilting of the oxygen octahedra by the $R_4^+$ and $M_3^+$ modes, the  $\Lambda_3$ mode has significant amplitude for both the oxygen and Pb atoms (see Appendix~D, Table~\ref{tab:Tab6}). 

The $\Lambda_3$ distortion is also found in certain Bi-containing perovskites, where it results in complex antiferroelectric displacements of Bi and O attributed to the stereochemical activity of the lone electron pair on the Bi$^{3+}$ cations \cite{Shi2013}. Given that lone pair instabilities are also observed in Pb$^{2+}$ -based perovskites it seems very likely that there are antiferroelectric-like distortions of the Pb and O in \Pb{}.
This is a particularly interesting observation because the  structural transition in \Pb{} is continuous and occurs in the metallic phase. Therefore, at temperatures above $T_{\rm N}$ \Pb{} can be classified as a `metallic antiferrolectric'. In 2013, the first example of a `metallic ferroelectric' was reported \cite{Shi2013} and attracted considerable attention because the electrostatic mechanisms that drive electric dipole formation and ferroelectric distortions are usually suppressed in metals due to screening by conduction electrons. The antiferroelectric state is also associated with a spatial charge separation and formation of well correlated electric dipoles, and also expected to be uncommon in metals. It would be of interest, therefore, to understand what drives the antiferroelectric-like structural distortion in \Pb{}. 

Despite being a metal, the magnetic susceptibility of \PbZn{} follows the Curie--Weiss law for temperatures above about 80\,K. The Weiss temperature of 214\,K indicates strong antiferromagnetic interactions between localized magnetic moments. However, we observe no sign of magnetic order in susceptibility or neutron diffraction down to 2\,K, indicating that the sample is strongly frustrated. At low temperatures the susceptibility deviates from the Curie--Weiss law. Possible reasons for this deviation include the presence of a small concentration of paramagnetic impurity,
and the build-up of short-range magnetic correlations. 
 
In a perfectly cubic double perovskite the Os atoms would form a face-centred cubic (FCC) lattice, which for AFM nearest-neighbor interactions is frustrated and has no unique magnetic ground state \cite{Kermarrec2015,Revelli2019}. As discussed above, \PbZn{} is distorted compared to the ideal cubic symmetry, and thus only approximately forms a FCC lattice but nevertheless remains significantly frustrated 
In \Pb{}, the magnetic frustration is nearly fully relieved, as evidenced by the onset of magnetic order at \Tc{} K, close to the Weiss temperature of 101\,K (which is only half that of \PbZn{}). The additional distortion away from the ideal FCC  lattice in \Pb{} could be what relieves the frustration.
It is not surprising that the lower symmetry, less frustrated structure has the greater tendency for magnetic order, but it has been noted previously that in other $5d^2$ double-perovskites it seems that decreasing symmetry (and increasing spin--orbit coupling) destabilizes magnetic order, the opposite of our results here \cite{Aczel2016}.

The average ordered magnetic moment in \Pb{} is only around 0.6\,$\mu_{\rm B}$/Os at 2\,K, and the entropy associated with the magnetic transition is $S_\text{mag}\approx 0.3R$,  consistent with such a small moment. 
A similar reduction in ordered moment was found in Sr$_2$MgOsO$_6$. \cite{Morrow2016} 
In that  material, density functional theory (DFT) calculations attributed the reduction in ordered moment to strong spin--orbit coupling, as well as to hybridization with the neighbouring oxygen atoms which carry up to 40\% of the moment. Such a hybridized moment would contribute to the susceptibility and heat capacity, but less so to neutron diffraction due to the magnetic form factor. Hybridization would explain the strong exchange interactions between the Os ions, and the broadening of the transitions in our RIXS data. 

A non-collinear magnetic order in \Pb{} would be quite surprising. Most other $5d^2$ double perovskites either have non-magnetic or spin-glass groundstates, or collinear ferro- or antiferromagnetic long range magnetic order \cite{Aczel2016}. The main difference between \Pb{} and these other double perovskites seems to be the structural distortion, which results in two crystallographically distinct Os sites. The splitting of the $t_{2g}$ levels is thus expected to be different for the two sites, which could contribute to the broadening of the peaks in our RIXS data, and might also result in different magnetic moments on the two sites. It is also possible that the average charge on the two different sites is different as reported for some perovskites\cite{Alonso1999}. 

We conclude by discussing the possible origins of the MIT in \Pb{}. The MIT coincides with the transition to magnetic order, indicating that the two phenomena are related. We observe anomalies in the lattice constants at the transition, but no change in the crystal symmetry to within the precision of our diffraction measurements. The low degree of frustration   makes it unlikely that the structural anomalies are driven by relief of frustration, in contrast to what is proposed for \PbCo{}.\cite{Princep2019}  On the other hand, the presence of magneto-elastic coupling indicates that the orbitals could play a role in the MIT, as has been suggested in frustrated multi-orbital $5d^3$ double perovskites \cite{Chen2018}. 

In some respects the MIT in \Pb{} resembles that in NaOsO$_3$, which was initially proposed to be a Slater transition \cite{LoVecchio2013}. In both compounds, a transition from a Curie--Weiss metal  to an insulator or bad metal occurs simultaneously with AFM order, and the ordered moment is much smaller than the paramagnetic moment. In NaOsO$_3$, the MIT  occurs over a temperature range of some 400\,K, and such a continuous transition has been argued to be a signature of a spin-driven Lifshitz transition, rather than a Slater transition\cite{Kim2016a}. The electronic configuration of Os in NaOsO$_3$ is $5d^3$ corresponding to half-filled $t_{2g}$ states, whereas it is $5d^2$ in \Pb{}. Hence, the effect of spin--orbit coupling is expected to be larger in \Pb{}, and as a result the mechanism of the MIT may be different from that in NaOsO$_3$.

The pyrochlore-structure $5d$ oxides Cd$_2$Os$_2$O$_7$ and $A_2$Ir$_2$O$_7$ also have metal--insulator transitions which coincide with magnetic order, but these cannot be Slater transitions because the magnetic order has propagation vector ${\bf k}=0$ which does not change the periodicity of the crystal  potential.\cite{Yamaura2012,Sagayama-PRB-2013,Donnerer-PRL-2016,Guo-PRB-2016,Jacobsen-PRB-2020}. The magnetic order in \Pb{}, on the other hand, has propagation  vector $\kkk=(\tfrac{1}{2},\tfrac{1}{2},0)$. This magnetic order results in a doubling in the periodicity of the crystal potential, consistent with one of the requirements of a Slater transition.\cite{Slater1951}  

\section{Conclusion}

Our study has revealed that \PbZn{} exhibits no phase transitions between 2 and 350\,K,  and remains a paramagnetic metal throughout this temperature range. By contrast, \Pb{} is found to have several unusual and interesting phases. It undergoes a structural transition at high temperatures (above 500\,K) to a distorted phase with antiferroelectric-like displacements of the Pb and O atoms despite having metallic conductivity. Therefore, it could be classified as an `antiferroelectric metal', in the same sense that LiOsO$_3$ (Ref.~\onlinecite{Shi2013}) is referred to as a `ferroelectric metal' (a term originally coined by Anderson and Blount \cite{Anderson-Blount}).  It also exhibits a metal--insulator transition (at 80\,K) coincident with the onset of antiferromagnetic order which doubles the periodicity of the crystal potential. Our data suggest that the transition is spin-driven, which makes it a candidate for a Slater-type MIT. Angle-resolved photoemission measurements combined with electronic structure calculations to determine the Fermi surface and locate the band gap in momentum space would be very valuable to constrain the mechanism of the MIT in \Pb{} with greater certainty.

\acknowledgments
This work was supported by the U.K. Engineering and Physical Sciences Research Council (grant numbers EP/N034872/1 and EP/N034694/1), and the National Natural Science Foundation of China (Grant No. 11874264). The work in Japan was supported in part by JSPS KAKENHI (grant numbers JP20H05276 and JP19H05819), a research grant from Nippon Sheet Glass Foundation for Materials Science and Engineering ($\sharp$40-37), and Innovative Science and Technology Initiative for Security (grant number JPJ004596), ATLA.  HJ acknowledges funding from the EU Horizon 2020 programme under the Marie Sklodowska-Curie grant agreement No 701647. Experiments at the ISIS Neutron and Muon Source were supported by beam-time allocations from the Science and Technology Facilities Council. We thank A. Senyshyn for help with data collection during a preliminary neutron diffraction experiment performed at the SPODI instrument operated by FRM II at the MLZ, Germany. We also acknowledge the Diamond Light Source for the provision of time on Beamline I11 under Proposal EE9839, and the European Synchrotron Radiation Facility for provision of time on ID20. We thank Roger Johnson and Mike Glazer for help in solving the nuclear and magnetic structure of \Pb{}, Nat Davies for assistance with data analysis, and Masao Arai, Franz Lang and Ping Zheng for  preliminary experiments and discussions.

Two authors (HJ and HLF) contributed equally to this work.

\newpage

\appendix

\section{Magnetic susceptibility of polycrystalline  Pb$_2$Os$_2$O$_7$.}

\begin{figure}
    \centering
    \includegraphics[width=0.48\textwidth]{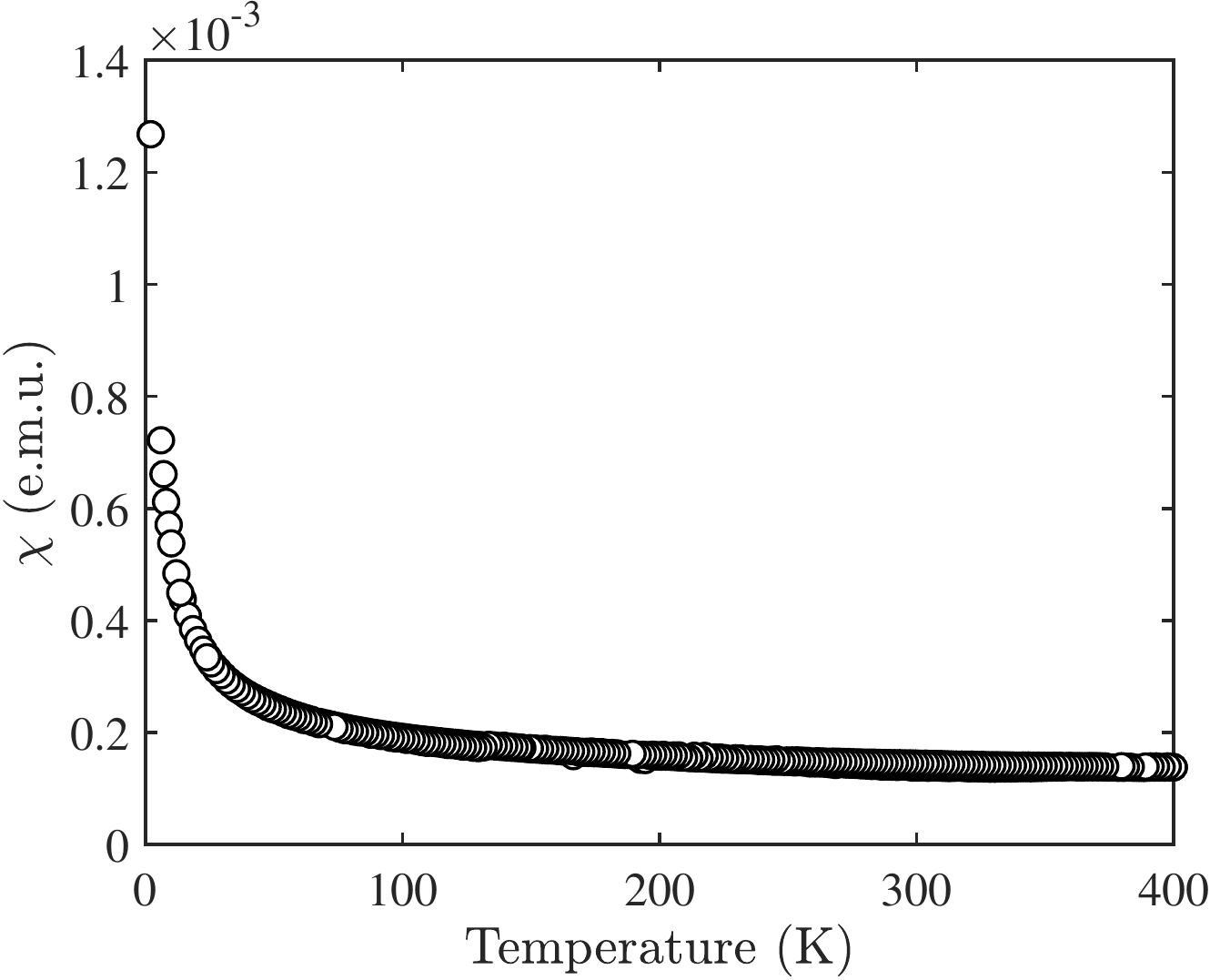}
    \caption{Magnetic susceptibility of the pyrochlore Pb$_2$Os$_2$O$_7$, showing paramagnetic behaviour down to 2 K.}
    \label{fig:Fig8}
\end{figure}

The temperature dependence of the magnetic susceptibility of Pb$_2$Os$_2$O$_7$ is shown in Fig.~\ref{fig:Fig8}. There is no evidence for any magnetic transitions in the measured temperature range between 2 and 400\,K.

\section{Structural parameters from neutron powder diffraction}

\begin{figure}
    \centering
    \includegraphics[width=0.48\textwidth]{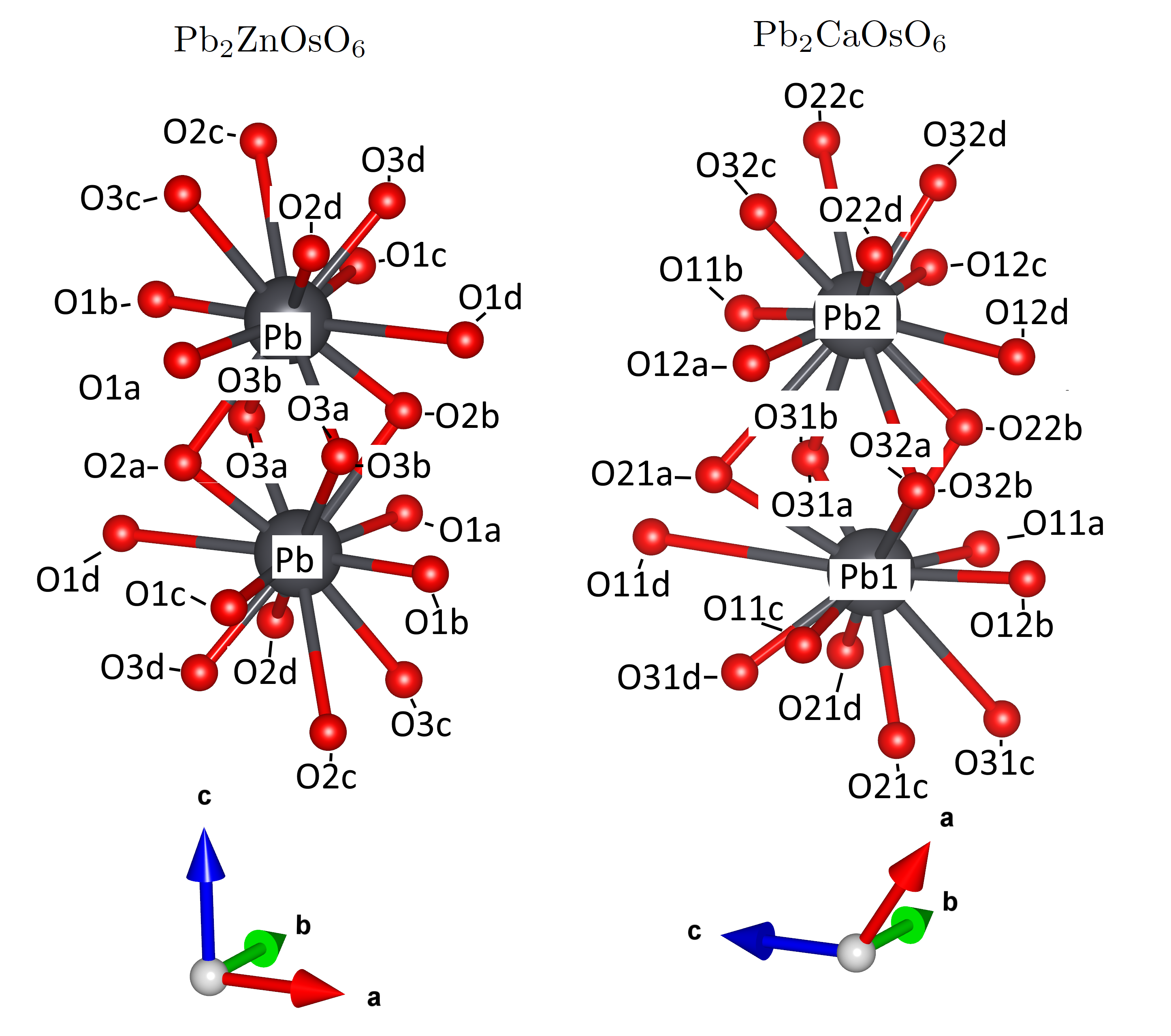}
    \caption{The local coordination of Pb in \PbZn{} (left) and \Pb{} (right). The labels of the oxygen atoms correspond with Table~\ref{tab:Tab6}.}
    \label{fig:Fig9}
\end{figure}

Structural parameters of \Pb{} and \PbZn{} are given in Table~\ref{tab:Tab2}. The coordination of the Os sites is discussed in the main text. As illustrated in Fig.~\ref{fig:Fig9}, the local environment of the Pb atoms is significantly distorted relative the cubic parent in which the 12 nearest oxygen atoms form a cuboctahedron with Pb at the center.  Table~\ref{tab:Tab6} gives the corresponding Pb--O distances for \PbZn{} and \PbCa{}. In \PbCa{}, the Pb--O distances deviate from the mean by up to 0.6\,\AA{}, and the Pb1 and Pb2 atoms are displaced from the center of the surrounding oxygens by 0.052\,\AA{} and 0.038\,\AA{}, respectively. In \PbZn{}, the deviation is up to 0.4\,\AA{}, and the displacement from the center is 0.019\,\AA{}.

\begin{table}[]
    \centering
\begin{tabular}{cccccc}
\hline \hline
Name &$x$ &$y$ &$z$ &$B_\text{iso}$ (\AA{})\\ \hline
Pb1 & 0.7561(6) & 0.5204(10) & 0.8814(5) & 0.68(8) \\ 
Pb2 & 0.7624(6) & 0.4913(10) & 0.3785(5) & 0.68(8) \\ 
Ca1 & 0.509(1) & -0.010(1) & 0.757(1) & 1.1(2) \\ 
Os1 & 0 & 0 & 0 & 0.55(9) \\ 
Os2 & 0 & 0 & 0.5 & 0.55(9) \\ 
O11 & 0.7787(10) & -0.069(1) & -0.0512(7) & 0.8(1) \\ 
O12 & 0.7684(9) & 0.037(1) & 0.3944(7) & 0.8(1) \\ 
O21 & 0.5296(9) & 0.225(2) & 0.6149(8) & 1.2(2) \\ 
O22 & 0.5274(10) & 0.338(1) & 0.1459(8) & 1.2(2) \\ 
O31 & 0.5850(9) & 0.699(2) & 0.6591(7) & 0.9(2) \\ 
O32 & 0.496(1) & 0.794(1) & 0.0802(8) & 0.9(2) \\ \hline \hline
\end{tabular}

\vspace{0.2cm}

\begin{tabular}{cccccc}
\hline \hline
Name &$x$ &$y$ &$z$ &$B_\text{iso}$ (\AA{})\\ \hline 
Pb1 & 0.0059(9) & 0.5153(9) & 0.247(2) & 1.29(6) \\ 
Zn1 & 0 & 0 & 0 & -0.3(2) \\ 
Os1 & 0 & 0 & 0.5 & 1.7(2) \\ 
O1 & -0.066(1) & -0.007(2) & 0.251(1) & 1.3(2) \\ 
O2 & 0.246(2) & 0.301(2) & 0.036(2) & 2.4(3) \\ 
O3 & 0.304(2) & 0.766(2) & 0.037(2) & 0.5(2) \\ \hline \hline 
\end{tabular}

    \caption{Refined structural parameters of \Pb{} and \PbZn{} at 2\,K from neutron powder diffraction data recorded on WISH. The Bragg R-factors are 4.68 and 8.91 for \Pb{} and \PbZn{}, respectively. The space group for both compounds is P2$_1$/n. The lattice parameters at $T = 2$\,K  are $a=10.0812(3)$\,\AA{}, $b=5.689(1)$\,\AA{}, $c=11.837(4)$\,\AA{}, $\beta=125.32(2)$ (\Pb{}), and  $a=5.6329(2)$\,\AA{}, $b=5.6059(2)$\,\AA{}, $c=7.9201(2)$\,\AA{}, $\beta=89.96(1)^\circ$ (\PbZn{}). }
    \label{tab:Tab2}
\end{table}

\begin{table}
\begin{tabular}{cccccc}
\hline \hline
          & \PbZn{}         &      & \Pb{}   &      &  \\ \hline
Atom      & Pb--O            & Atom & Pb1--O     & Atom & Pb2--O     \\ \hline
O1a       & 2.710(11)       & O11a & 2.434(11) & O12a & 3.107(10) \\
O1b       & 2.481(9)        & O12b & 2.805(13) & O11b & 2.334(13) \\
O1c       & 2.955(11)       & O11c & 3.425(11) & O12c & 2.591(10) \\
O1d       & 3.158(9)        & O11d & 3.752(12) & O12d & 3.072(12) \\
O2a       & 2.461(17)       & O21a & 3.108(9)  & O22a & 2.533(8)  \\
O2b       & 2.8468(18)      & O22b & 2.803(13) & O21b & 3.397(13) \\
O2c       & 3.244(17)       & O21c & 2.895(9)  & O22c & 3.298(9)  \\
O2d       & 2.735(16)       & O21d & 2.432(13) & O22d & 3.027(13) \\
O3a       & 2.748(15)       & O31a & 2.385(9)  & O32a & 3.416(9)  \\
O3b       & 3.255(17)       & O32b & 3.349(14) & O31b & 3.441(12) \\
O3c       & 2.842(18)       & O31c & 3.426(8)  & O32c & 2.499(9)  \\
O3d       & 2.457(17)       & O31d & 2.649(12) & O32d & 2.459(13) \\ \hline \hline
\end{tabular}
\caption{The local coordination of the Pb atoms in \PbZn{} and \Pb{}, showing the Pb--O distance in \AA{} for the 12 nearest O atoms. }
\label{tab:Tab6}
\end{table}

\section{Irreducible representations}
\begin{table}[]
    \centering
    \begin{tabular}{c c c c c }
    \hline \hline
         IRs& $\psi$ &  Component & Atom 1 & Atom 2 \\ \hline
         $\Gamma_2^3$   & $\psi_1$ &Real      & (1 0 0) & (0 0 0)\\
                        &          &Imaginary & (0 0 0) & (1 0 0) \\
                        & $\psi_2$ &Real      & (0 1 0) & (0 0 0)\\
                        &          &Imaginary & (0 0 0)  &(0 -1 0) \\
                        & $\psi_3$ &Real      & (0 0 1) & (0 0 0)\\
                        &          &Imaginary & (0 0 0)  & (0 0 1) \\
        $\Gamma_4^3$    & $\psi_1$ &Real      & (1 0 0) & (0 0 0)\\
                        &          &Imaginary & (0 0 0) & (-1 0 0) \\
                        & $\psi_2$ &Real      & (0 1 0) & (0 0 0)\\
                        &          &Imaginary & (0 0 0)  & (0 1 0) \\
                        & $\psi_3$ &Real      & (0 0 1) & (0 0 0)\\
                        &          &Imaginary & (0 0 0)  &(0 0 -1) \\ \hline \hline  \end{tabular}
    \caption{The irreducible representations and basis vectors for the Os1 and Os2 sites in \Pb{} with space group P$2_1$/n and $\kkk=(\tfrac{1}{2},\tfrac{1}{2},0)$ obtained from representational analysis using BASIREPS. For the Os1 site, atom 1 refers to the position $0,0,0$ and Atom 2 to $\tfrac{1}{2},\tfrac{1}{2},\tfrac{1}{2}$. For the Os2 site, atom 1 is at position $0,0,\frac{1}{2}$ and atom 2 is at $\frac{1}{2},\frac{1}{2},0$.}
    \label{tab:Tab6}
\end{table}

For the space group P2$_1$/n and $\kkk=(\tfrac{1}{2},\tfrac{1}{2},0)$, the magnetic  representation  $\Gamma_\text{mag}$ for the two Os sites (Wyckoff positions 2a and 2c) decomposes into the irreducible representations
\begin{align}
    \Gamma_\text{mag} = \Gamma_2^3 + \Gamma_4^3.
\end{align}

Both representations are three-dimensional and complex, and the corresponding basis functions are summarized in Table \ref{tab:Tab6}. The basis functions that transform as $\Gamma_2$ and $\Gamma_4$ are complex conjugates and must be combined to obtain real moments on the Os sites.

\section{Decomposition of structures}
In Tables \ref{tab:Tab6} and \ref{tab:Tab6} we show the decomposition of the \PbZn{} and \Pb{} structures into the symmetrized displacive modes of the parent cubic Pm$\bar{3}$m structure. 
\begin{table*}
\begin{tabular}{llll} \hline \hline
Irrep $(k)$ & Order parameter & Site irrep & Amplitude  \\ \hline
 & Strains &  &  \\ \hline
$\Gamma_{1}^+(0,0,0)$ & $(a),e_{xx}+e_{yy}+e_{zz}$ & $(a)$ & -0.00001\\
$\Gamma_{3}^+(0,0,0)$ & $(a,0),-e_{xx}-e_{yy}+2e_{zz}$ & $(a)$ & -0.00277\\
$\Gamma_{5}^+(0,0,0)$ & $(a,b,b),e_{yz}$ & $(a)$ & 0.00337\\
$\Gamma_{5}^+(0,0,0)$ & $(a,b,b),e_{xy}+e_{xz}$ & $(b)$ &  0.00056\\
\hline & Pb displacement & &  \\ \hline
$R_5^+$ $(1/2,1/2,1/2)$ & $(a,b,b)$ & $T_{lu}(a)$ & -0.00318\\ 
 &  & $T_{lu}(b)$ & -0.03256\\ 
$X_5^+$ $(0,1/2,0)$ & $(0,0;0,0;a,a)$ & $T_{lu}(a)$ & 0.201\\ \hline & O displacement & &  \\ \hline
$R_1^+$ $(1/2,1/2,1/2)$ & $(a)$ & $A_{2u}(a)$ & 0.20079\\ 
$R_3^+$ $(1/2,1/2,1/2)$ & $(a,0)$ & $A_{2u}(a)$ & -0.04084\\ 
$R_4^+$ $(1/2,1/2,1/2)$ & $(0,a,-a)$ & $E_u(a)$ & 1.1013\\ 
$R_5^+$ $(1/2,1/2,1/2)$ & $(a,b,b)$ & $E_u(a)$ & 0.03593\\ 
 &  & $E_u(b)$ & -0.00556\\ 
$X_5^+$ $(0,1/2,0)$ & $(0,0;0,0;a,a)$ & $E_u(a)$ & 0.11566\\ 
$M_2^+$ $(1/2,1/2,0)$ & $(a;0;0)$ & $A_{2u}(a)$ & -0.00112\\ 
$M_3^+$ $(1/2,1/2,0)$ & $(a;0;0)$ & $E_u(a)$ & 0.69843\\ 
$M_5^+$ $(1/2,1/2,0)$ & $(a,a;0,0;0,0)$ & $E_u(a)$ & 0.03369\\ 
\hline\hline
\end{tabular}
\caption{Decomposition of the structure of \PbZn{} into the symmetrized displacive modes of the parent cubic $Pm\bar{3}m$ perovskite structure ($a=3.97$\, \AA{}, Pb $1b$ (1/2,1/2,1/2), Zn/Os $1a$ (0,0,0), O $3d$ (1/2,0,0)). The column "Irrep $(k)$" shows the irreducible representations of the   $Pm\bar{3}m$ space group and the arms of the wave vector star involved. The column "Order parameter" lists the projections of the reducible order parameter onto the corresponding irreducible subspace (the same symbol in different positions indicates equal order parameter components). The column "Site irrep" shows the corresponding point-group symmetry irrep of the local Wyckoff position, and the order parameter component in brackets. The column "Amplitude" displays the amplitude of the displacive modes in \AA{}. }
\label{tab:Tab6}
\end{table*}

\begin{table*}
\begin{tabular}{llll} \hline \hline
Irrep $(k)$ & Order parameter & Site irrep & Amplitude  \\ \hline
 & Strains &  &  \\ \hline
$\Gamma_{1}^+(0,0,0)$ & $(a),e_{xx}+e_{yy}+e_{zz}$ & $(a)$ & 0.00024\\
$\Gamma_{3}^+(0,0,0)$ & $(a,\sqrt{3}a),2e_{xx}-1e_{yy}-1e_{zz}$ & $(a)$ & 0.00182\\
$\Gamma_{5}^+(0,0,0)$ & $(a,b,-a),e_{xz}-1e_{yz}$ & $(a)$ & 0.00783\\
$\Gamma_{5}^+(0,0,0)$ & $(a,b,-a),e_{xy}$ & $(b)$ & -0.02793\\
\hline & Pb displacement & &  \\ \hline
$\Lambda_1$ $(1/4,1/4,1/4)$ & $(0,0;0,0;0,0;a,0)$ & $T_{lu}(a)$ & 0.12338\\ 
$\Lambda_3$ $(1/4,1/4,1/4)$ & $(0,0,0,0;0,0,0,0;a,-a,-1/\sqrt{3}a,-1/\sqrt{3}a;b,-b,\sqrt{3}b,\sqrt{3}b)$ & $T_{lu}(a)$ & -0.23899\\ 
 &  & $T_{lu}(b)$ & -0.00237\\ 
$R_5^+$ $(1/2,1/2,1/2)$ & $(a,b,-a)$ & $T_{lu}(a)$ & -0.01068\\ 
 &  & $T_{lu}(b)$ & -0.21487\\ 
$X_5^+$ $(0,1/2,0)$ & $(0,0;a,-a;0,0)$ & $T_{lu}(a)$ & -0.09609\\ 
\hline & Ca displacement & &  \\ \hline
$\Lambda_1$ $(1/4,1/4,1/4)$ & $(0,0;0,0;0,0;a,0)$ & $T_{lu}(a)$ & -0.03793\\ 
$\Lambda_3$ $(1/4,1/4,1/4)$ & $(0,0,0,0;0,0,0,0;a,-a,-1/\sqrt{3}a,-1/\sqrt{3}a;b,-b,\sqrt{3}b,\sqrt{3}b)$ & $T_{lu}(a)$ & -0.11614\\ 
 &  & $T_{lu}(b)$ & -0.15423\\ 
\hline & O displacement & &  \\ \hline
$\Lambda_1$ $(1/4,1/4,1/4)$ & $(0,0;0,0;0,0;a,0)$ & $A_{2u}(a)$ & 0.00671\\ 
 &  & $E_u(a)$ & -0.00284\\ 
$\Lambda_2$ $(1/4,1/4,1/4)$ & $(0,0;0,0;0,a;0,0)$ & $E_u(a)$ & -0.10574\\ 
$\Lambda_3$ $(1/4,1/4,1/4)$ & $(0,0,0,0;0,0,0,0;a,-a,-1/\sqrt{3}a,-1/\sqrt{3}a;b,-b,\sqrt{3}b,\sqrt{3}b)$ & $A_{2u}(a)$ & -0.30797\\ 
 &  & $A_{2u}(b)$ & -0.15126\\ 
 &  & $E_u(a)$ & -0.99142\\ 
 &  & $E_u(b)$ & -0.49071\\ 
 &  & $E_u(a)$ & 1.4942\\ 
 &  & $E_u(b)$ & 0.98874\\ 
$R_1^+$ $(1/2,1/2,1/2)$ & $(a)$ & $A_{2u}(a)$ & -1.0125\\ 
$R_3^+$ $(1/2,1/2,1/2)$ & $(a,\sqrt{3}a)$ & $A_{2u}(a)$ & 0.04599\\ 
$R_4^+$ $(1/2,1/2,1/2)$ & $(a,0,a)$ & $E_u(a)$ & 1.6103\\ 
$R_5^+$ $(1/2,1/2,1/2)$ & $(a,b,-a)$ & $E_u(a)$ & 0.00755\\ 
 &  & $E_u(b)$ & 0.27841\\ 
$X_5^+$ $(0,1/2,0)$ & $(0,0;a,-a;0,0)$ & $E_u(a)$ & -0.2628\\ 
$M_2^+$ $(1/2,1/2,0)$ & $(0;0;a)$ & $A_{2u}(a)$ & -0.04517\\ 
$M_3^+$ $(1/2,1/2,0)$ & $(0;0;a)$ & $E_u(a)$ & 0.50508\\ 
$M_5^+$ $(1/2,1/2,0)$ & $(0,0;0,0;a,-a)$ & $E_u(a)$ & -0.1971\\ 
\hline\hline
\end{tabular}
\caption{Decomposition of the structure of \Pb{} into the symmetrized displacive modes of the parent cubic $Pm\bar{3}m$ perovskite structure ($a=4.10631$\, \AA{}, Pb $1b$ (1/2,1/2,1/2), Ca/Os $1a$ (0,0,0), O $3d$ (1/2,0,0)). The column "Irrep $(k)$" shows the irreducible representations of the   $Pm\bar{3}m$ space group and the arms of the wave vector star involved. The column "Order parameter" lists the projections of the reducible order parameter onto the corresponding irreducible subspace (the same symbol in different positions indicates equal order parameter components). The column "Site irrep" shows the corresponding point-group symmetry irrep of the local Wyckoff position, and the order parameter component in brackets. The column "Amplitude" displays the amplitude of the displacive modes in \AA{}. }
\label{tab:Tab6}
\end{table*}

\FloatBarrier

\end{document}